\documentclass[runningheads]{llncs}

\usepackage{graphicx}
\usepackage{amsmath}
\usepackage{amsfonts}
\usepackage{amssymb}
\usepackage{amsbsy}
\usepackage{xcolor}
\usepackage{bm}
\usepackage{times}
\usepackage{comment}
\usepackage{mymacros}
\usepackage[style=base]{caption}
\usepackage{subcaption}

\begin{document}

\title{Removing Operational Friction Using Process Mining}
\subtitle{Challenges Provided by the Internet of Production (IoP)}
%
%
\author{Wil M.P. van der Aalst\inst{1,2} \and
Tobias Brockhoff \inst{1} \and
Anahita Farhang Ghahfarokhi \inst{1} \and
Mahsa Pourbafrani \inst{1} \and
Merih Seran Uysal \inst{1} \and
Sebastiaan J. van Zelst\inst{1,2}
 }
\authorrunning{W. van der Aalst et al.}
%
\institute{
Process and Data Science (Informatik 9), RWTH Aachen University, Aachen, Germany \and
Fraunhofer-Institut f\"{u}r Angewandte Informationstechnik, Sankt Augustin, Germany \\
\email{wvdaalst@rwth-aachen.de}}

\maketitle              
\begin{abstract}
Operational processes in production, logistics, material handling, maintenance, etc., are supported by cyber-physical systems combining hardware and software components.
As a result, the digital and the physical world are closely aligned, and it is possible to track operational processes in detail (e.g., using sensors).
The abundance of event data generated by today's operational processes provides
opportunities and challenges for process mining techniques
supporting process discovery, performance analysis, and conformance checking.
Using existing process mining tools, it is already possible to automatically discover process models and 
uncover performance and compliance problems.
In the DFG-funded Cluster of Excellence  ``Internet of Production'' (IoP), 
process mining is used to create ``digital shadows'' to improve a wide variety of operational processes.
However, operational processes are dynamic, distributed, and complex.
Driven by the challenges identified in the IoP cluster, we work on novel techniques for  
\emph{comparative} process mining 
(comparing process variants for different products at different locations at different times),
\emph{object-centric} process mining
(to handle processes involving different types of objects that interact), and
\emph{forward-looking} process mining (to explore ``What if?'' questions).
By addressing these challenges, we aim to develop valuable ``digital shadows'' that can be used to remove operational friction.

\end{abstract}

\section{Introduction}
\label{sec:intro}

Data are collected about anything, at any time, and at any place.
Of course, operational processes in production, logistics, material handling, and maintenance are no exception.
In \cite{DBLP:books/sp/Aalst16} the term \emph{Internet of Events} (IoE) was introduced to reflect that
machines, products, vehicles, customers, workers, and organizations are increasingly connected to the internet 
and that our capabilities to track and monitor these entities advanced in a spectacular manner. 
Notions such as the \emph{Internet of Things} (i.e., all physical objects that are able to 
connect to each other and exchange data) and \emph{Industry 4.0} 
(i.e., the fourth industrial revolution enabled by smart interconnected machines and processes),
reflect our abilities to monitor operational processes.

The DFG-funded Cluster of Excellence ``Internet of Production'' (IoP) joins the efforts of over 200 engineers and computer scientists with the goal to fundamentally transform the way industrial production takes place by using data-driven techniques (\url{www.iop.rwth-aachen.de}). A key concept within IoP is the notion of a \emph{digital shadow}, i.e.,
automatically generated models that can be used to control and improve production processes.
Although IoP also considers novel production technologies and materials science and engineering, 
the lion's share of IoP activities is devoted to improving production processes.
This position paper focuses on \emph{operational processes} that involve \emph{physical objects} 
(products, machines, etc.) and have \emph{discrete} process steps.
One of the main goals of IoP is to remove \emph{operational friction} in such processes.
Much like physical friction, operational friction refers to phenomena that consume additional effort and energy and lead to less-than-optimal results. Examples of operational friction are unnecessary rework, delays, waste, deviations, fraud, missing products, unresponsiveness, and communication loops.

\emph{Process mining} can be used to remove operational friction by making conformance and performance problems 
visible in such a way that it becomes clear what the root-causes of such problems are \cite{DBLP:books/sp/Aalst16}.
Starting point for process mining are the event data mentioned before.
Events may take place inside 
a production facility, inside a transportation system, inside a product, or
inside an enterprise information system. 
Events may be triggered by people, machines, or organizations.
\emph{Process discovery} techniques use event data to create process models 
describing the operational processes in terms of their key activities.
These process models reveal the actual processes and can be extended to show bottlenecks and outlier behavior.
\emph{Conformance checking} techniques compare observed behavior (i.e., event data)
with modeled behavior (i.e., process models).
These techniques can be used to show deviations, i.e., behaviors different from what is expected or desired.
Process models may also include probabilities, time distributions, and business rules.
Therefore, process mining also includes a range of techniques enabling \emph{predictive and prescriptive analytics}.

The process models created and used by process mining techniques form the \emph{digital shadows} that are intended to manage, control, and improve operational processes. The notion of a digital shadow is at the core of IoP.
The notion of a digital shadow is closely related to the notion of a \emph{digital twin} \cite{caise2020-jarke}.
Digital twins are complex models providing digital counterparts to physical production artifacts 
(i.e., the physical twins), e.g., detailed simulation models that update and change as their physical counterparts change.
The digital shadow concept is broader and includes the traces of the actual production processes, 
i.e., \emph{digital shadows based on process mining include both process models and actual event data}.
For example, when analyzing a bottleneck, it is possible to use the process model to drill-down into event data that caused the problem. Moreover, the process model and event data can be used to predict the trajectory of 
current process instances and anticipate the effect of interventions. 
This makes process mining a key ingredient of IoP.

The process mining discipline emerged around 20 years ago \cite{DBLP:books/sp/Aalst16},
and today there are over 35 commercial process mining tools (e.g., Celonis, Disco, UiPath/ProcessGold, myInvenio, PAFnow, Minit, QPR, Mehrwerk, Puzzledata, LanaLabs, StereoLogic, Everflow, TimelinePI, Signavio, and Logpickr).
Many of the larger organizations (especially in Europe) have adopted this technology.
For example, within Siemens over 6000 employees are using Celonis Process Mining to remove operational friction and increase automation. 
Despite the widespread adoption of process mining and the availability of easy-to-use commercial tools,
the process mining discipline is relatively young and there are still many open challenges.
In this position paper, we focus on three questions particularly relevant for IoP:
\begin{itemize}
  \item \emph{How to compare different process variants (over time, over locations, over different case types)?}
  The same production process may be performed at different locations. 
  It is valuable to understand why operational friction is less at some of these locations.
  Also, the performance may not be constant over time and we may see drifts, e.g., bottlenecks are shifting.
  Therefore, we need to support comparative process mining \cite{DBLP:conf/apbpm/Aalst13,hcse2014keynote,DBLP:conf/simpda/BoltLAG15,DBLP:conf/simpda/HompesBADB15}
  \item \emph{How to deal with processes involving different interacting objects?}
  Traditionally, each event refers to a single case (e.g., an order). 
  This is made possible by flattening event data. This may lead to convergence and divergence problems.
  The same event may be replicated for multiple cases or unrelated events may appear to be related.
  This leads to misleading diagnostics. Moreover, depending on the question at hand,
  the event data need to be extracted differently from systems such as SAP, Oracle, and Microsoft Dynamics. 
  Therefore, we need to support \emph{object-centric} process mining techniques that are able to handle different types of objects (i.e., case notions) in one model \cite{DBLP:conf/sefm/Aalst19}.
  \item \emph{How to improve the process using forward-looking techniques?}
  The initial focus of process mining was on diagnosing historical event data.
  However, the ultimate goal is to improve processes and remove operational friction.
  Therefore, we need process mining techniques that are \emph{forward-looking} and that provide \emph{actionable} results, e.g., automated alerts, interventions, reconfigurations, policy changes, and redesign.
  To anticipate the effect of process interventions, a tighter integration with simulation is needed \cite{DBLP:conf/caise/Aalst10,DBLP:conf/scsc/Aalst18,DBLP:conf/otm/PourbafraniZA19,sim-YAWL-ProM-dke}.
\end{itemize}

The remainder of this position paper is organized as follows.
Section~\ref{sec:pm} introduces an overview of process mining techniques.
Section~\ref{sec:re} introduces a running example based on which we discuss the potentials and challenges of applying process mining to operational processes from a research and process owner's perspective.
Section~\ref{sec:extpm} uses the running example to illustrate the capabilities of existing process mining tools.
Section~\ref{sec:cpm} discusses the need for comparative process mining and presents some of the initial capabilities to support process cubes, concept drift, and process comparison using advanced concepts such as the earth mover's distance.
Section~\ref{sec:oopm} presents the problems related to convergence and divergence in processes due to events that refer to multiple objects.
Initial support for object-centric process mining techniques based on process cubes is presented.
In Section~\ref{sec:fpm}, we advocate a shift from backward-looking process mining to more forward-looking forms of analysis combining simulation and process mining. Next to traditional discrete event simulation 
with models generated from event data, we also need simulation approaches operating at a higher abstraction level.
For this purpose, we propose to combine process mining and system dynamics.
Section~\ref{sec:concl} concludes the paper.

\section{Process Mining: Removing Friction in Operational Processes}
\label{sec:pm}

To help organization to remove operational friction, process mining reveals unnecessary rework, delays, waste, deviations, fraud, missing products, lost responses, etc.
This is not so easy.  As an example, take the Order-to-Cash (O2C) process of a large multinational that processes over 30 million ordered items per year. 
These 30 million cases (i.e., instances of the O2C process) generate over 300 million events per year. Over 60 different activities may occur. 
Although the O2C process is fairly standard, over 900,000 process variants can be observed in one year! 
These variants describe different ways of executing this process. 
In such processes, often 80\% of all cases are described by just 20\% of the process variants.
However, the remaining 20\% of the cases generate 80\% of the process variants and are often responsible for a disproportional part of an organization's operational friction.
Such problems cannot be tackled by traditional process modeling or workshops, because these focus on the dominant 20\% of the process variants.
Process mining aims to provide insights into real-life operational processes using the event data at hand.
\begin{figure}[htb]
{
\centering
\includegraphics[width=0.80\textwidth]{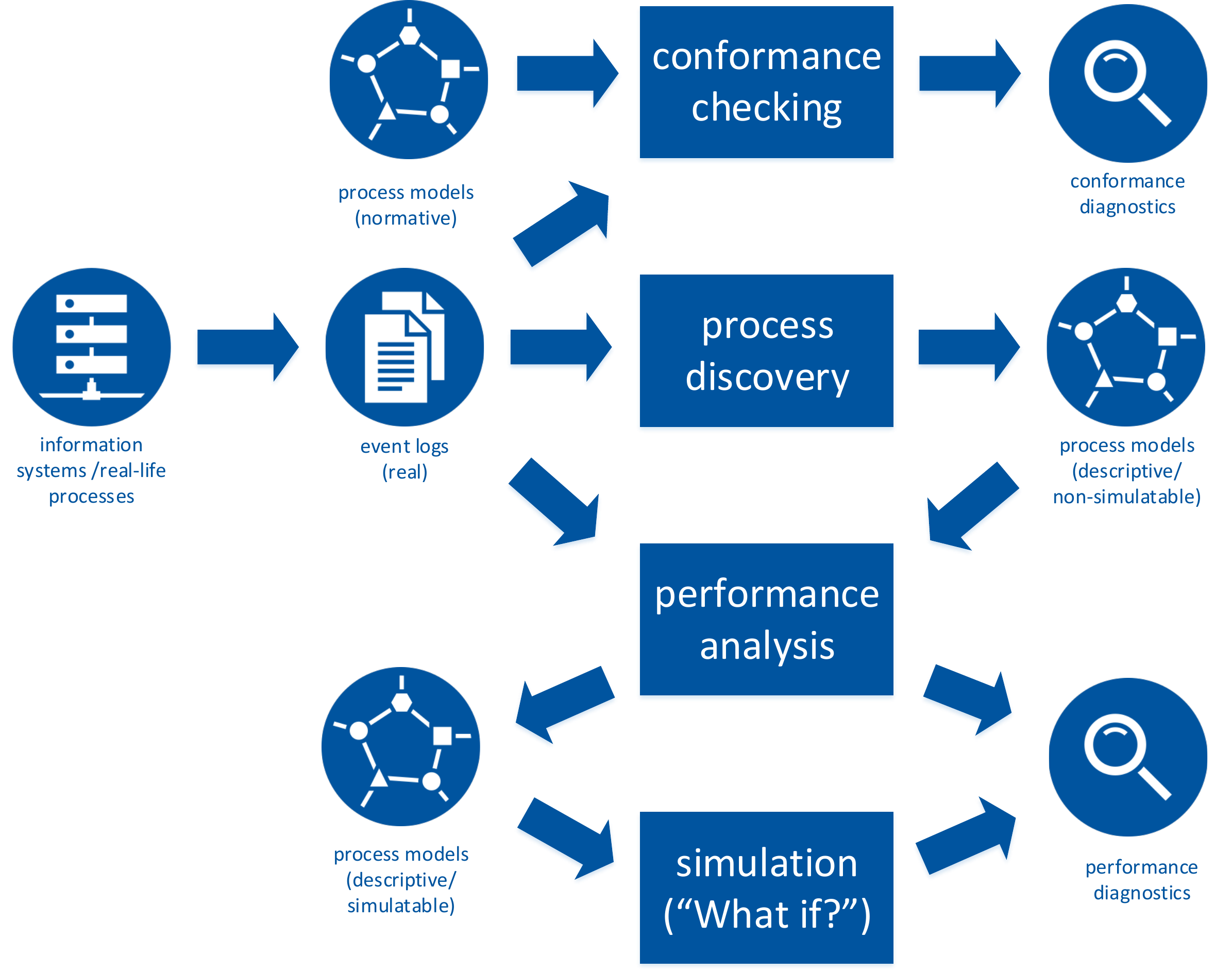}
\caption{An overview of process mining showing how event logs, process models, and diagnostics relate.
}\label{fig-pm-types}
}
\end{figure}

The starting point for traditional process mining is an \emph{event log} extracted from the information system(s).
Each \emph{event} in such a log refers to an \emph{activity}
possibly executed by a \emph{resource} at a particular \emph{time} and for a particular \emph{case} \cite{DBLP:books/sp/Aalst16}.
An event may have many more attributes, e.g., transactional information (e.g., start, complete, and abort), costs, customer, location, and unit.
Typical challenges are \emph{data quality} and finding a \emph{suitable case notion}.
A typical data quality problem is the granularity of the timestamps in the event log. 
Some events may only have a date (e.g., 29-9-2020) and are manually recorded, whereas other 
events may have a millisecond precision (e.g., 29-9-2020:15.36.45.567) and are captured fully automatic.
As a result, the precise ordering is unknown or uncertain.
Events are grouped into traces using the selected case notion.
However, as detailed in Section~\ref{sec:oopm}, this is not always easy.
One event may refer to one order, ten parts, two machines, one person, and a location.
Different events may share objects and one event may refer to many objects.

As Figure~\ref{fig-pm-types} shows, event data are used for process discovery, conformance analysis, and performance analysis \cite{DBLP:books/sp/Aalst16}.
\emph{Process discovery} techniques automatically learn process models based on an event log.
A process model can be represented using different modeling languages, e.g.,
Petri nets, BPMN models, UML activity diagrams, DFGs, automata, and process trees.
Such a process model specifies how a case can be handled. The simplest model is a so-called Directly Follows Graph (DFG).
The nodes in such a model are the activities and two special nodes: the start and end of the process.
The arcs describe how the process transitions from one node to another. Arcs can be decorated with frequencies and mean durations.
For example, 500 times activity ``send invoice'' was followed by activity ``make payment'' and the average time between both activities was 2.45 days.
DFGs are simple and can be constructed efficiently, also for huge events logs. However, DFGs fail to capture concurrency and often lead to spaghetti-like underfitting models.
Petri nets, BPMN models, UML activity diagrams, and process trees are able to express concurrency and there are many process discovery algorithms to discover such models.
Here, we abstract from the actual representation of process models and focus on the more general concepts.

Conformance checking techniques take as input both a process model and an event log.
The event log is replayed on the model to uncover discrepancies between the observed and modeled behavior.
It may be the case that an activity occurs although this is not possible according to the model,
or an activity should have occurred according to the model, but did not happen in reality.
It is also possible to see that behavior allowed by the process model never happens.
If the process model is extended with temporal or resource constraints, conformance checking may reveal that something takes too long or is performed by the wrong person.

By replaying the event data on the model, it is also possible to show performance problems. 
Since events have timestamps, it is possible to measure delays, next to routing probabilities.
Even when the event data do not completely fit the model, it is possible to ``squeeze'' the data into to model using alignment computations. 
Process models with performance information can be used to analyze bottlenecks and other performance problems.
Moreover, by combining stochastic models learned from historic event data with event data from running cases, it is possible to predict performance and compliance problems.
\begin{figure}[htb]
{
\centering
\includegraphics[width=0.99\textwidth]{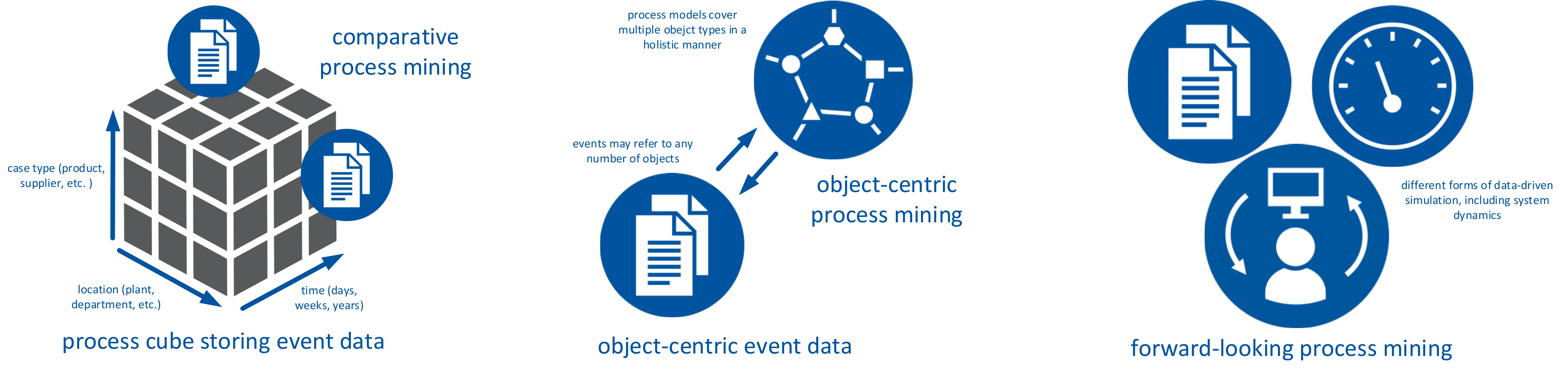}
\caption{Different research directions relevant for the improvement of operational processes: 
\emph{comparative} process mining, \emph{object-centric} process mining, and \emph{forward-looking} process mining.
}\label{fig-challenges}
}
\end{figure}

Although existing process mining techniques and tools have proven to be a powerful means to remove operational friction,
several challenges remain. In the context of our Cluster of Excellence ``Internet of Production'' (IoP), 
we address the three challenges already mentioned in the introduction.
Figure~\ref{fig-challenges} sketches these challenges and corresponding solution directions.
To compare different process variants (over time, over locations, over different case types) 
we use process cubes and novel techniques to compare subsets of events (see Section~\ref{sec:cpm}).
Process cubes store event data using any number of dimensions.
Novel techniques based on the Earth Mover's Distance (EMD) notion are used to compare two different event logs (or a log and model) while also considering time and resources  (also detailed in Section~\ref{sec:cpm}).
To deal with processes involving different interacting objects, we use object-centric event logs and corresponding process mining techniques (detailed in Section~\ref{sec:oopm}).
Traditional process mining techniques require the selection of a single case notion and each event refers to precisely one case.
However, for the types of processes we have in mind, this is often too restrictive.
Each event may refer to any number of objects and the case notion is just a view on the process.
Objects may refer to physical products (raw materials, end products, intermediate products, etc.), resources, or information.
Resources include machines, people, space, etc.
To make process mining more forward-looking, we support different types of simulation.
This allows us to anticipate the effects of process interventions and use process models as highly informative digital shadows.
Next to traditional Discrete Event Simulation (DES), we also support System Dynamics (SD) techniques that view the process at a higher aggregation level (see Section~\ref{sec:fpm}).
For SD models, we do not simulate the individual events, but use aggregated steps representing different time periods (e.g., days or weeks).

\section{Running Example}
\label{sec:re}

\begin{figure}[t!]
{
\centering
\includegraphics[width=\textwidth]{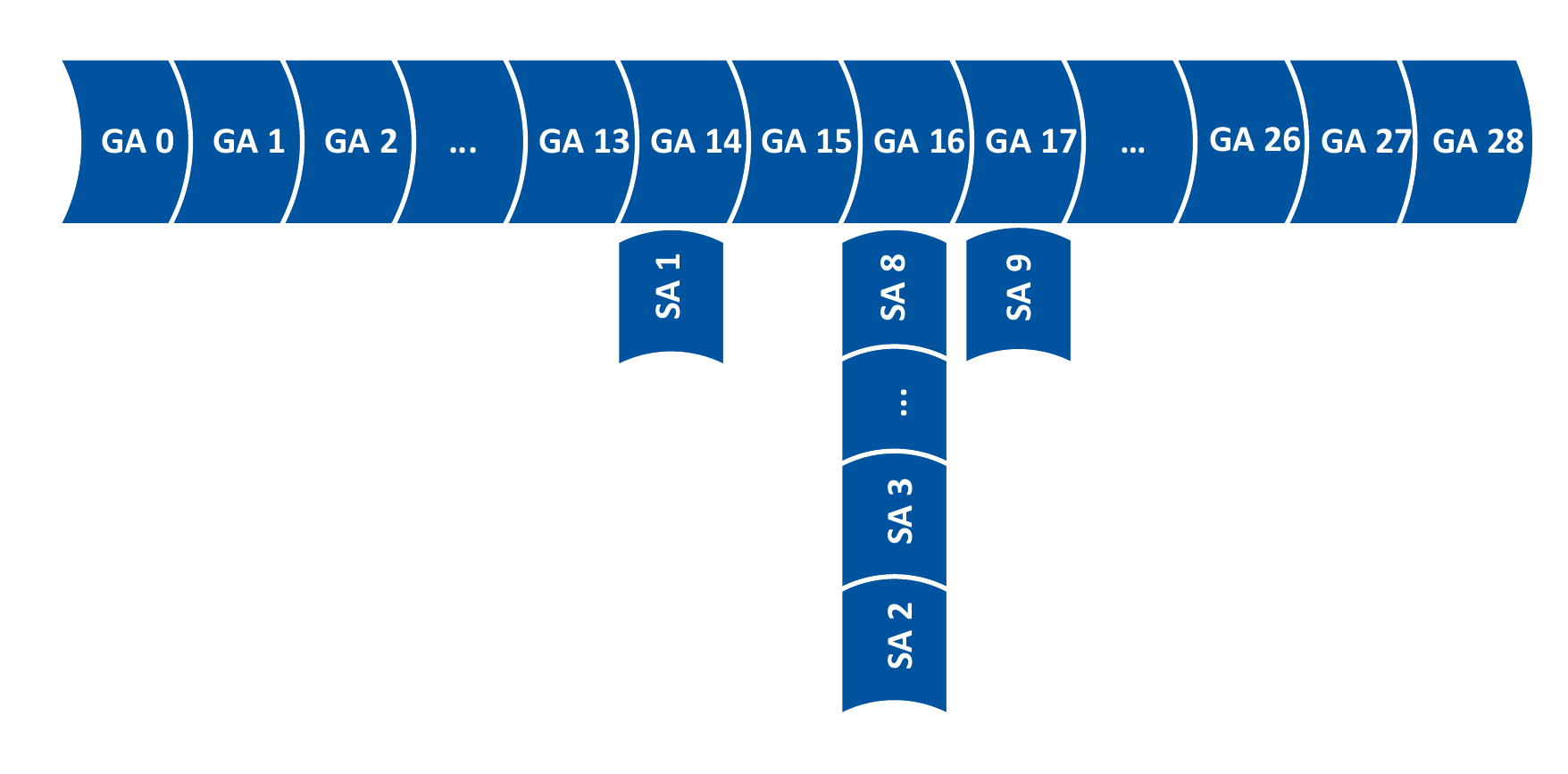}
\caption{A part of the simulated model of the production line, used for designing the simulation model. The production line includes one general assembly line with 28 stations and 33 sub-assembly stations.}\label{fig-base-model}
}
\end{figure}

Various types of operational processes can be found in industry. Automotive production lines are a common type of production lines that are used to execute operational processes. Automotive production lines are highly structured and activities therein are tightly coupled. In a simple production line, multiple assembly lines are coordinated in order to join in a single major assembly line. Due to the variability of activities in the automotive production lines and the strong dependencies between them, they are suitable for process mining analysis. Therefore, in order to illustrate process mining techniques for operational processes, this section introduces an extendable automotive production line model. The model is based on a production line at e.GO Mobile AG, a local car manufacturer located in Aachen. Due to reasons of confidentiality, we resort to simulated data and present our results without utilizing the real data of e.Go. A fragment of the simulated model is shown in Figure~\ref{fig-base-model}. It shows the production line is built around a sequence of general assembly stations where some stations depend on additional sequential sub-assembly stations. For example, the station \emph{SA9} is a prerequisite for the general assembly station \emph{GA17}. Each station can have at most one car at a time, during which the human operators execute specific tasks before the car proceeds to the next station. A car can only move along the general assembly line if the subsequent station is empty. We consider 28 stations in the general assembly line and 33 stations in the sub-assembly line. Several activities are executed at one station and most of the time the order of activities at the stations does not matter.
The production line is active between 8:00-17:00 o'clock and 5 days a week. 
In the following sections we demonstrate how process mining is able to help a company to steer and evolve their operational process. 
To this end, we extend the basic model in order to illustrate the challenges mentioned in the introduction. 
In Section~\ref{sec:cpm}, we present \emph{comparative process mining} techniques to answer the following questions:
\begin{itemize}
  \item How to compare performance-related factors of two different factories in two different locations? 
  \item How to investigate resource allocation in two different factories?
\end{itemize}
In Section~\ref{sec:oopm}, we discuss \emph{object-centric process mining} approaches to answer following questions:
\begin{itemize}
  \item How to deal with real-life data of the car factory extracted from information systems such as ERP systems in process mining?
  \item How to address challenges arising from having multiple interacting objects, e.g., order, products, and customers in processes? 
 \end{itemize}
In Section~\ref{sec:fpm}, we present \emph{forward-looking process mining} methods to answer following questions:
\begin{itemize}
  \item How to increase the monthly production rate of cars?
  \item Does a temporary queuing station for the cars help to increase the overall production rate?
\end{itemize}

Whereas the emphasis of this paper is on advanced process mining techniques, 
we first present a few common process mining techniques, e.g., basic approaches for process discovery and conformance checking. 
These examples diagnostics also help us to obtain an overview of the process used throughout this paper.

\section{Applying Existing Process Mining Techniques}
\label{sec:extpm}

Process mining provides a wide variety of techniques for process analysis \cite{DBLP:books/sp/Aalst16}. Examples of process mining techniques include process discovery algorithms that are able to find process models that describe the event log, conformance checking algorithms that compare process models with the event logs, and model enhancement algorithms that enrich the process model with additional information extracted from the event log. To gain more insights about the process, we apply standard process mining techniques 
(see \cite{DBLP:books/sp/Aalst16} for a comprehensive introduction). First, in Section~\ref{sec:explore}, we perform an explorative analysis by means of process visualization techniques, i.e., the dotted chart. In Section~\ref{sec:pd}, we discover a process model of the generated data. In Section~\ref{sec:cc}, we highlight deviations between the discovered process model and the event log by applying conformance checking techniques. In Section~\ref{sec:pa}, we address the performance problems in the production line through bottleneck analysis.

\subsection{Exploring Event Data}
\label{sec:explore}

\begin{figure}[h!]
{
\centering
\includegraphics[width=\textwidth]{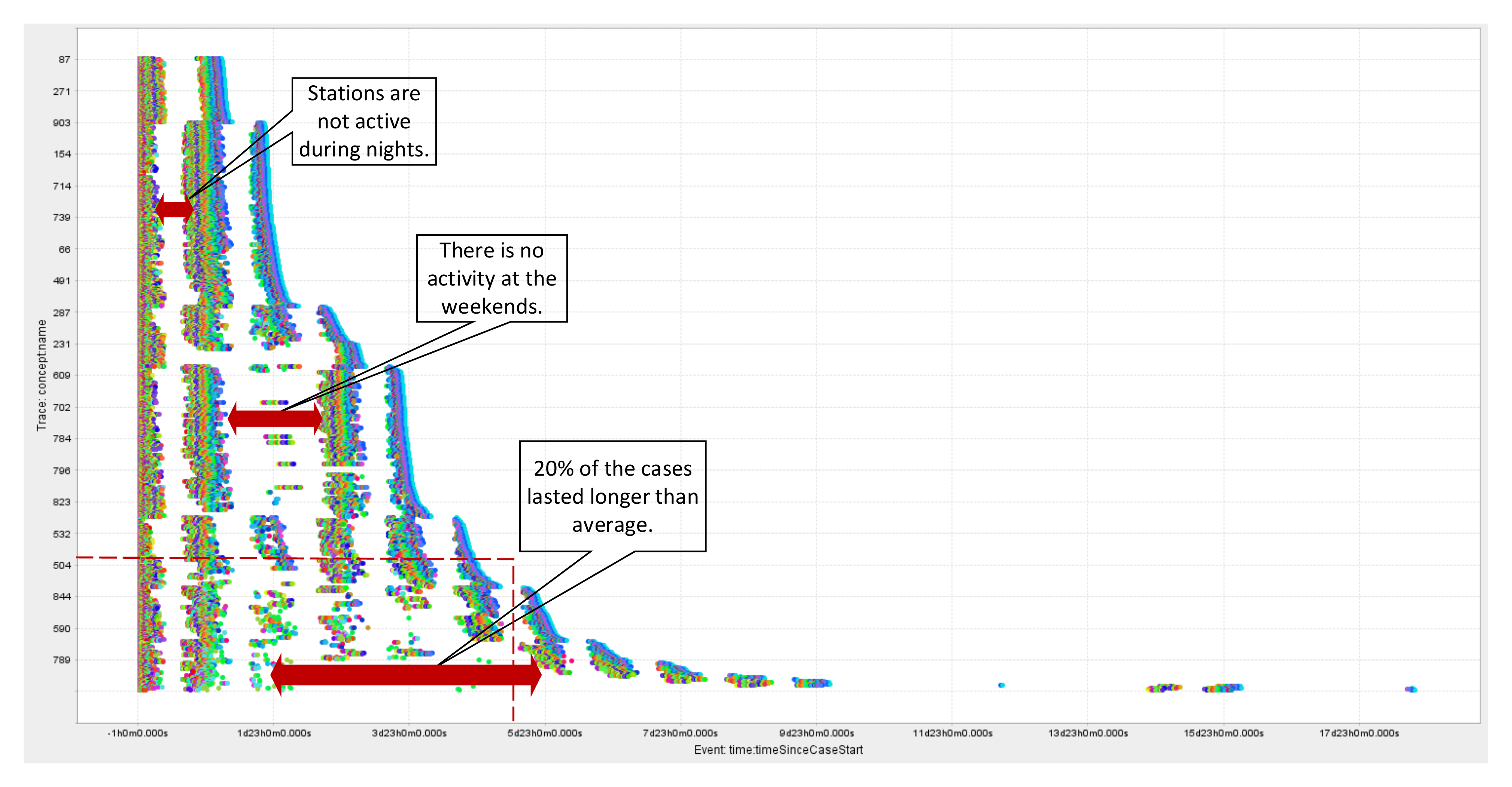}
\caption{Dotted chart for the generated event log. Working hours and working days are demonstrated in the dotted chart. Besides, we observe a big gap between events in 20\% of the cases, lasted longer than normal.}\label{fig-pm-dotted-chart}
}
\end{figure}

Explorative process analysis is a common starting point in order to obtain an overview of the execution of the process. A widely adopted tool that interactively visualizes multiple perspectives of an event log is the dotted chart~\cite{minseok_wits}. Figure~\ref{fig-pm-dotted-chart} presents the dotted chart for the generated data where each event is shown by a dot, colored according to the station it originated from. The x-axis and y-axis show time and the car id respectively. There are different ways of sorting the events in the dotted chart. However, in this dotted chart, events are sorted based on the case duration. As indicated in the figure, there are gaps between events that show the period in which no event occurs such as nights and weekends. We can observe 80/20 rule, also known as
the Pareto principle in the dotted chart. Approximately 20\% of the cases last exceptionally long. Following the Pareto principle, 80\% of the performance problems (e.g., delay in the production line) are caused by 20\% of the cases. Clearly, these 20\% of all the process instances are interesting cases for further analysis. 

Even though process visualization by means of a dotted chart helps the analysis, most of the analysis by using the dotted chart is empirical and not quantified. 
Thus, other process mining techniques such as process discovery and conformance checking techniques are required to find actionable results for improving operational processes.

\subsection{Process Discovery}
\label{sec:pd}

Although the visualization approach, discussed in Section~\ref{sec:explore}, provides an overview of the event data, it does not show the relationship between different activities.
Process discovery techniques take an event log as an input and generate a process model that captures the relation between activities and abstracts the observed behavior in the event log.

\begin{figure}[h!]
\centering
\includegraphics[ width=\textwidth]{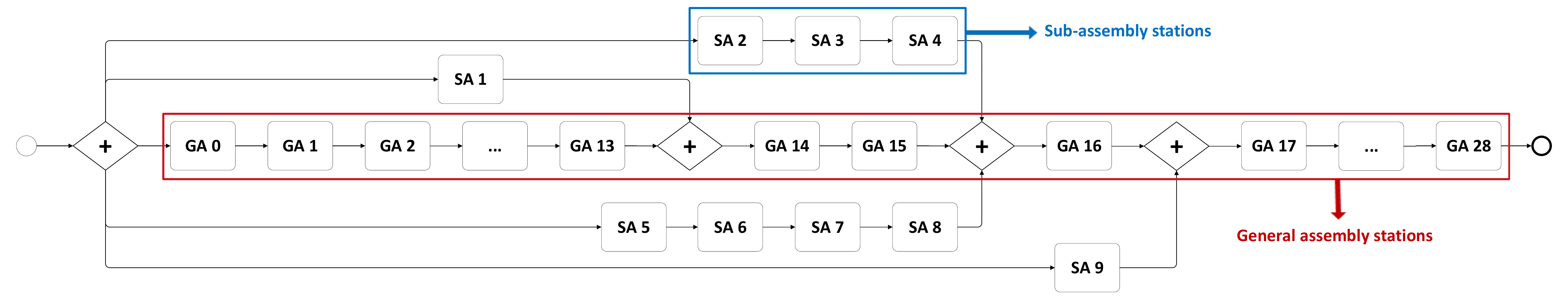}
\caption{An example process model of the production line using BPMN notation~\cite{DBLP:journals/csi/ChinosiT12}. The flow of activities for cars during the production line is shown. Activities at general assembly stations are always executed in a sequence. Also, activities at sub-assembly stations, required for a specific general assembly station, are executed sequentially.}\label{fig-processdiscovery}
\end{figure}

There are various techniques for process discovery, e.g., the Alpha-algorithm~\cite{DBLP:journals/tkde/AalstWM04}, the Inductive Miner~\cite{DBLP:conf/apn/LeemansFA14}, Region-based approaches~\cite{DBLP:books/sp/Aalst16}, and the Heuristics Miner~\cite{aal_min_icae}. For the resulting models, there are multiple different notations, e.g., Petri nets~\cite{Badouel-Darondeau-book-regions-2015}, BPMN~\cite{DBLP:journals/csi/ChinosiT12}, YAWL~\cite{YAWLbook}, UML~\cite{aalxrlorg}, etc. Figure~\ref{fig-processdiscovery} shows a fragment of the BPMN model for the example log by applying Inductive Miner. As the model shows, the general assembly stations are sequential and some of them have prerequisites, i.e., sub-assembly stations. For example, we start triggering the execution of activities at the station \emph{GA14}, when execution of activities at the stations \emph{SA1} and \emph{GA13} have been finished. Another aspect, captured by BPMN model, is concurrency. While the execution of the activities at the stations \emph{GA0}, \emph{GA1}, ..., and \emph{GA13} is sequential, they are all in parallel with the execution of the activities at the station \emph{SA1}. Therefore, in addition to the sequential relations, this discovery reveals the concurrency in activity executions. Note that in this discovery, we focused on the control-flow perspective. However, process discovery is not limited to the control flow perspective and it can also be applied to other perspectives of the process, such as the organizational perspective~\cite{DBLP:books/sp/Aalst16}. 

\subsection{Conformance Checking}
\label{sec:cc}

\begin{figure}[t]
\centering
\includegraphics[ width=\textwidth]{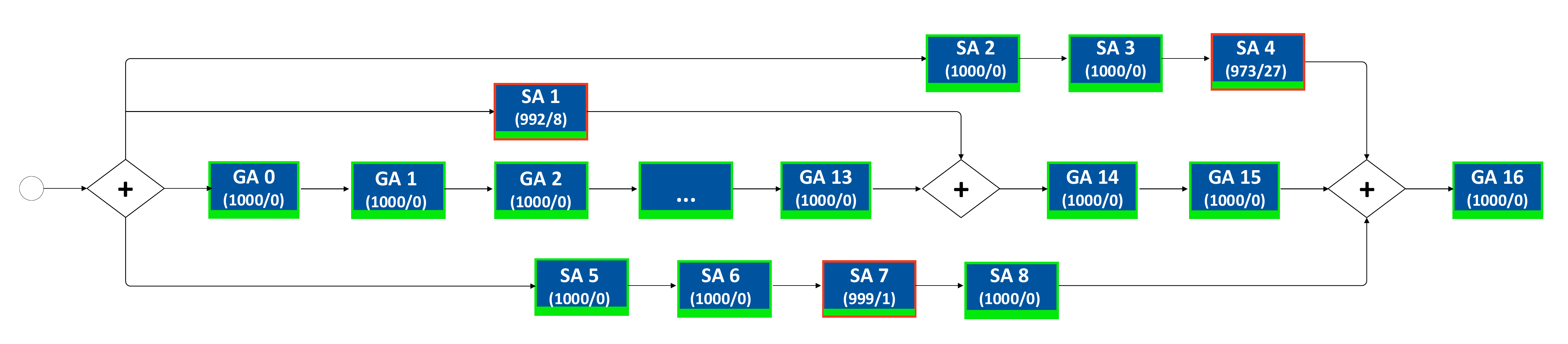}
\caption{A part of the detected deviations in the process. Activities with red square show deviations while activities with green square show that there was no deviation in that activity.}\label{fig-model-deviation}
\end{figure}

Given a process model, conformance checking assesses the similarity of the behavior described by the model and the real behavior as recorded in the event log. Therefore, conformance checking techniques~\cite{DBLP:conf/bpm/AdriansyahDA10} are used to identify deviations of the production plan. 
Since the event data used originate from an idealized simulation model, 
there are no deviations in the initial setting. However, we manipulate the simulation model such that for some stations deliberately deviations are injected. One of the common techniques, used in conformance checking, are alignments. In alignments, we check the synchronization between the behavior in the event log and behavior according to the model~\cite{DBLP:journals/widm/AalstAD12}.
An example of alignments is depicted in Figure~\ref{fig-model-deviation} which shows a fragment of the mined BPMN model with the results of conformance checking. As this figure shows, using conformance checking, deviations of the example process from the previously mined model are easily detected and visualized. Two numbers in parenthesis, below the name of the station, clarify the number of deviations. For instance, for 27 cars, activities at the station \emph{SA4} have not been executed before activities at the station \emph{GA16} and after the activities at the station \emph{SA3}. These cars have skipped the station \emph{SA4} and they may cause quality problems after the production. Accordingly, identifying these types of deviations in the production line can help the production line managers to control the quality and steps of the process accurately. 

\begin{figure}[t]
\centering
\includegraphics[ width=\textwidth]{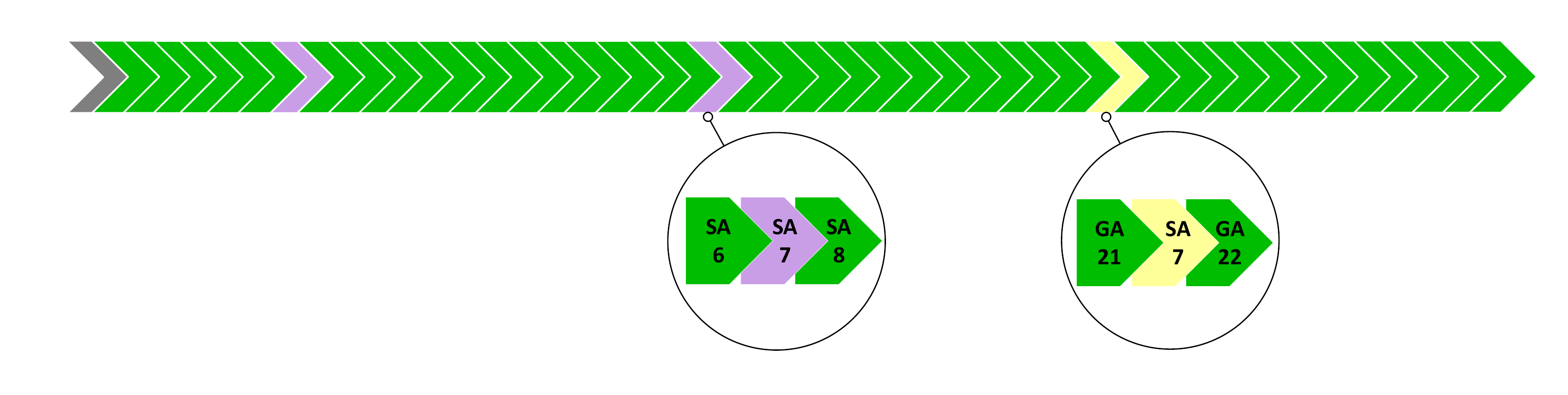}
\caption{An alignment for a specific car. A car ideally passes 61 stations. However, there are mismatches between the discovered model and the event log, captured in the alignment by different colors.}\label{fig-alignment}
\end{figure}

Figure~\ref{fig-alignment} shows the results of conformance checking using the alignments-based technique. In alignments, we check for a car, whether the behavior of the model and the behavior of the events related to that car match with each other or not. We have two types of misalignment which are model move and log move. In model move, a move in the log cannot be mimicked by a move in the model. In Figure~\ref{fig-alignment}, there are two model moves (purple color). For example, based on the model, we need the station \emph{SA7} in the event log between the stations \emph{SA6} and \emph{SA8}. However, it is missed in this case. In log move, a move in the model cannot be mimicked by a move in the log. In Figure~\ref{fig-alignment}, there is a log move (yellow color). Based on the event log, this car moves from the station \emph{GA21} to the station \emph{SA7} and from the station \emph{SA7} to the station \emph{GA22}. However, it cannot happen according to the model. Typically, by capturing such misstatements, we can discover the deviations that affect the quality of the final product. However, in order to increase the performance of the production line we need to analyze the performance of the stations to find the bottlenecks.

\subsection{Performance Analysis}
\label{sec:pa}

\begin{figure}[t]
\centering
\includegraphics[ width=\textwidth]{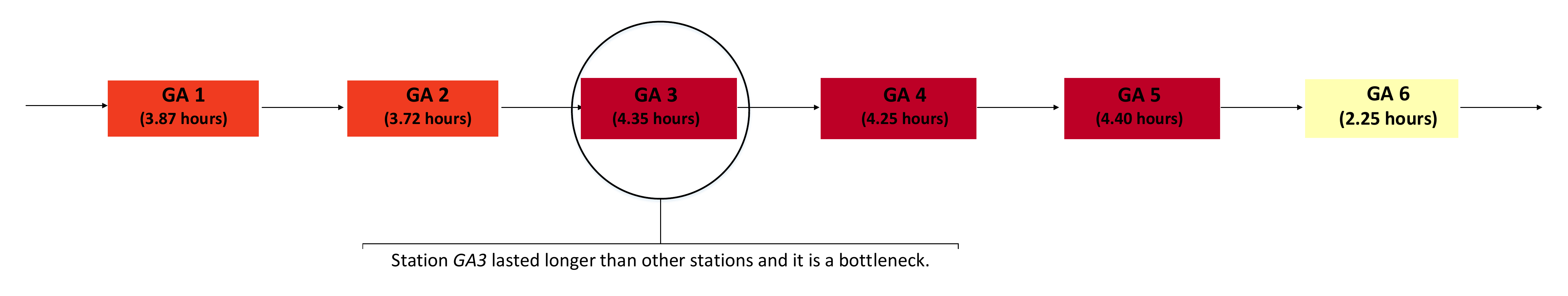}
\caption{Snapshot of the performance view of the general assembly line. The average service times of the stations are shown.}\label{fig-bottleneck}
\end{figure}

The performance of production lines is one of the important metrics which needs to be observed and improved continuously~\cite{DBLP:journals/widm/AalstAD12}. After verifying the suitability of the given model using conformance checking, this section assesses the performance of the process by means of bottleneck analysis. In Figure~\ref{fig-bottleneck}, we depict a projection of performance diagnostics, on top of the discovered process model. Within the figure, the darker colors of the stations show performance problems and bottlenecks in the process. For instance, the dark red color of the stations \emph{GA3}, \emph{GA4}, and \emph{GA5} indicates the long service times at these stations (i.e., the time actually worked on the car). Most of the time, improving the performance of bottlenecks such as the station \emph{GA6} improves the overall performance of the process. In general, process mining bottleneck analysis reveals this kind of problems and subsequent actions can be taken accordingly. In this performance analysis, we considered service time of the stations. However, it is possible to analyze the performance based on the waiting times for the different stations (i.e., the time that a car is waiting for the station to become available).

\begin{figure}[t]
\centering
\includegraphics[width=\textwidth]{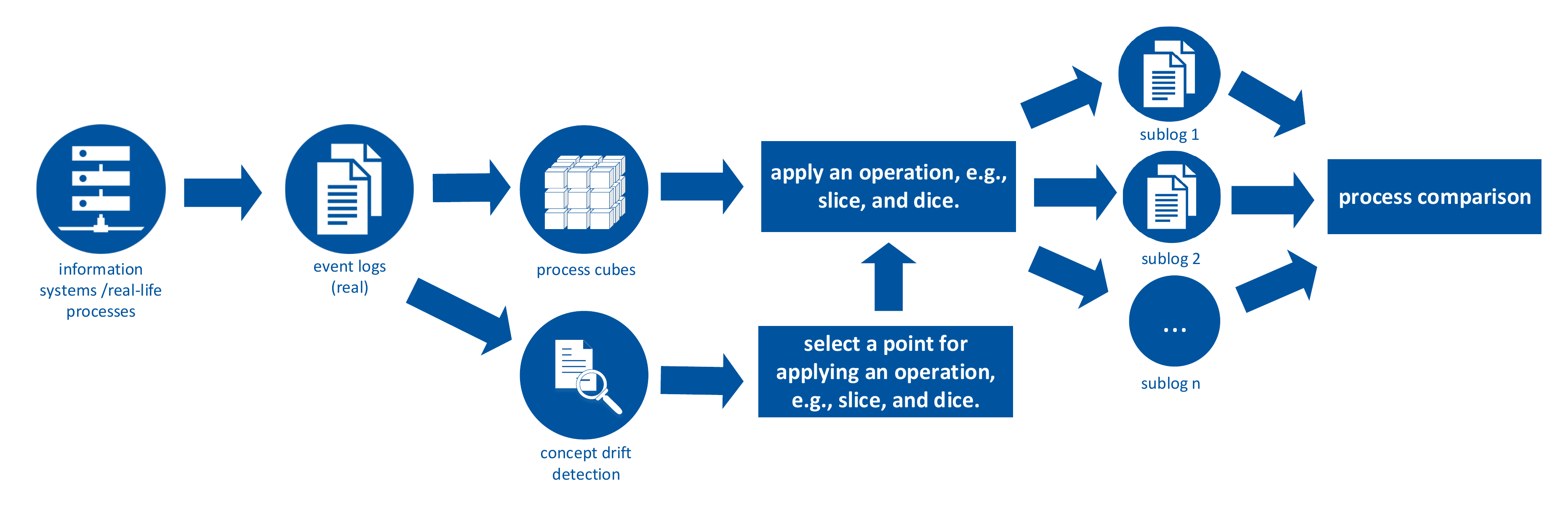}
\caption{Process cubes are built on the event logs extracted from information systems. We detect the change point in the regular patterns of the process by concept drift detection. Therefore, based on the change point, we apply process cube operations such as \emph{slice}. The output of such process cube operations are sublogs, used for process comparison.}\label{fig-comparison}
\end{figure}

\section{Comparative Process Mining}
\label{sec:cpm}

The techniques presented in Section~\ref{sec:extpm} capture the current state of the process using the complete event log. In this analysis, we do not consider variability in the processes which can be derived from the heterogeneity of demand in different seasons. In order to isolate process variations, process comparison, which systematically investigates the presence and absence of systematic differences, has recently gained interest. The application is not limited to a single process instance but can also consider multiple instances of similar processes. For example, suppose that we have two car factories in two different locations, implementing the production lines described in Section~\ref{sec:re}. Comparing performance-related metrics of these two factories with each other or even comparing the performance metrics of a single company in different time windows of a year is valuable from the business perspective. Characteristics related to the performance, such as the duration of the underlying production process for the comparison of the two factories, are of importance in operational management. This comparison can be addressed by splitting event data into process variants by using existing process mining approaches in a way that differences between variants are exposed. To provide better insights into current approaches for process comparison, we first show how event data can be organized considering different dimensions of variability, which enables basic process comparison in Section~\ref{sec:processcube}. While the former method assumes prior knowledge about the expected variability, Section~\ref{sec:drift} assesses how this can naturally be complemented by process change detection approaches and the emerging challenges for operational processes. Finally, in Section~\ref{sec:comp}, given the methods to organize the data, we introduce a challenging process comparison scenario, 
which motivates further research on performance-oriented operational process comparison.

\subsection{Process Cubes}
\label{sec:processcube}

\begin{figure}[t]
\centering
\includegraphics[width=0.9\textwidth, height = 0.28\textheight]{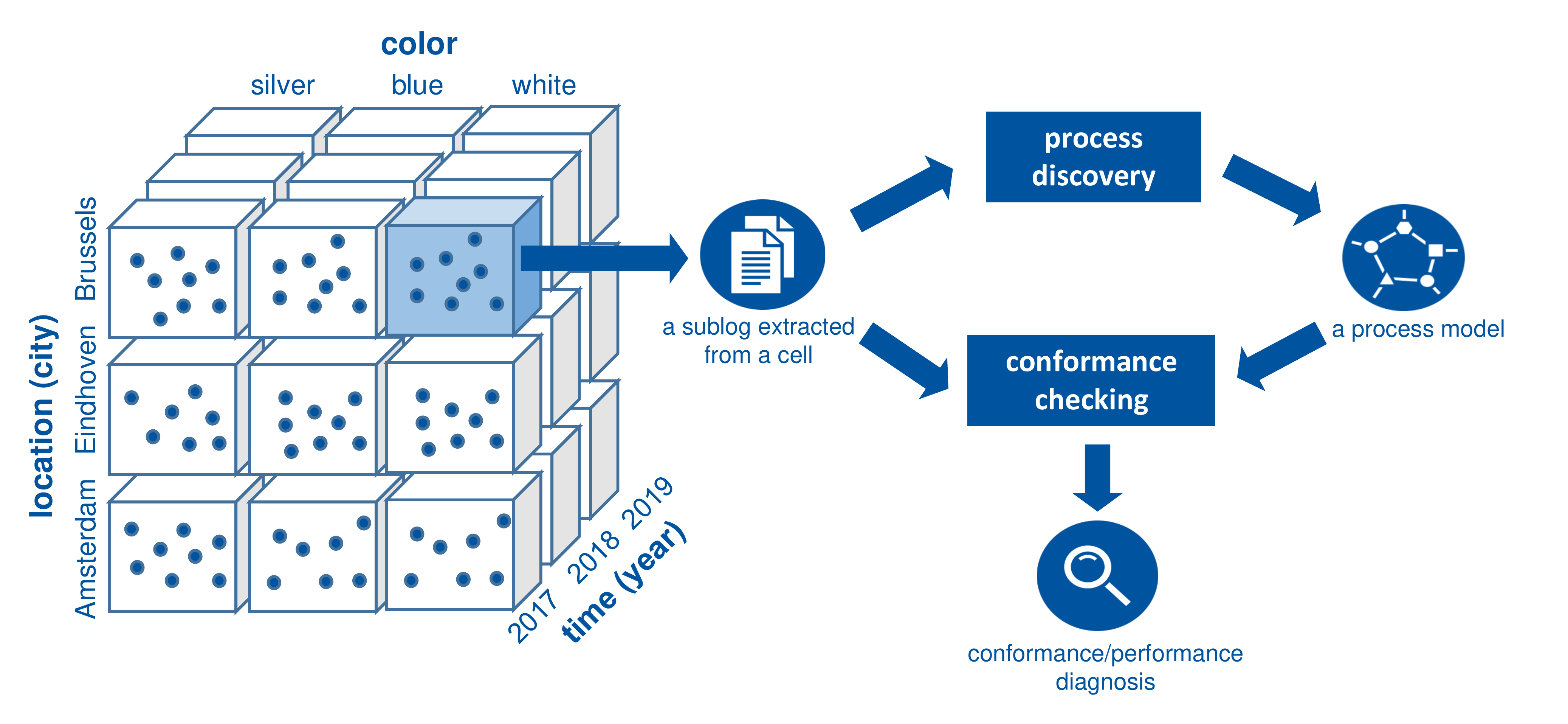}
\caption{An example of a three-dimensional process cube. The selected cell contains all events related to white cars produced in Brussels in \emph{2017}.}\label{fig-cube-cell}
\end{figure}

Classical process mining techniques focus on analyzing a process through processing its corresponding event log as a whole. However, there may exist variability in the process. To consider this variability in the process, we use process cubes which isolate the different processes. As shown in Figure~\ref{fig-cube-cell}, a process cube consists of multiple dimensions that refer to the properties of the event log, e.g., time, color, and location. Each cell in this process cube refers to all events related to the cars with a specific color, in a specific region, and in a particular time window. To gain more insights about the specific cell, process mining techniques, e.g., process discovery and conformance checking techniques can be applied to a collection of events extracted from the cells of the process cube~\cite{DBLP:conf/simpda/BoltLAG15}.

\begin{figure}[tb]
\centering
\includegraphics[width=1\textwidth, height = 0.50\textheight]{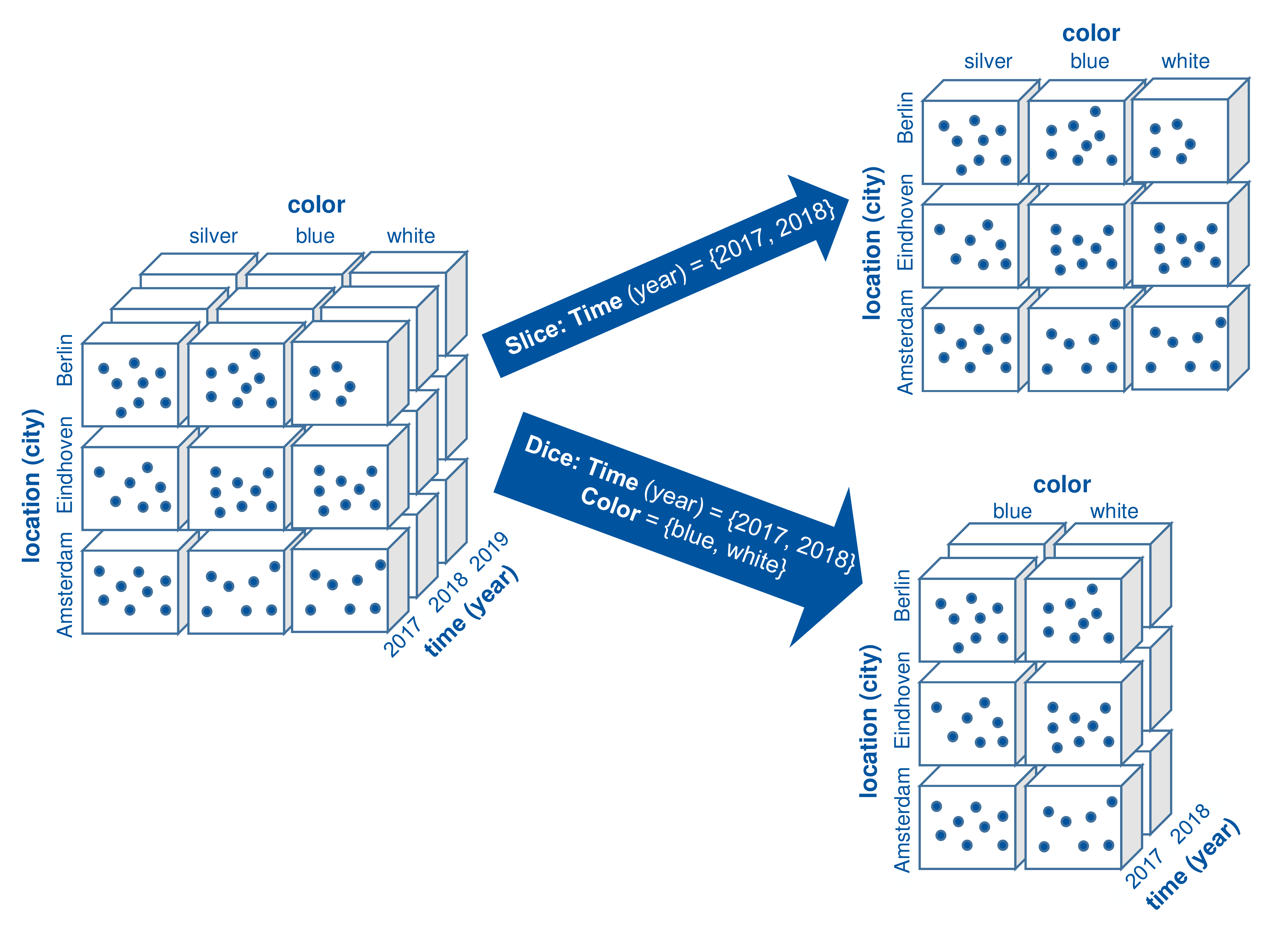}
\caption{Example of \emph{slicing} and \emph{dicing} operations in process cubes. A \emph{slice} of the cube for the \emph{time} dimension and \emph{dice} of the cube for the \emph{time} and \emph{color} dimensions are shown.}\label{fig-cube-slice-dice}
\end{figure}

The notion of process cubes is inspired by application of Online Analytical Processing (OLAP) in process mining~\cite{event-cube-coopis2011}. OLAP is a well-known concept in Business Intelligence (BI) that facilitates data discovery and management reporting. First, OLAP considers simple numerical analysis, e.g., plotting car production duration in a factory against the months of the year. In~\cite{chen2009graph,li2007mining,DBLP:conf/icde/LiuRGGWAM10}, OLAP operations, i.e., \emph{slice}, \emph{dice}, \emph{drill-down}, and \emph{roll-up} were also applied to non-numerical data. Although the idea of the process cube is related to OLAP, there are significant differences between OLAP and process cubes. We can not use OLAP for process data directly, because, for example, as a design choice in a process cube, cells are associated with both process models and event data. In a process cube, event data and process models are directly related. We can discover models from event data extracted from cells of the cube and compare observed and modeled behavior~\cite{DBLP:conf/apbpm/Aalst13}. Different implementations of process cubes are provided in~\cite{DBLP:conf/caise/BoltA15,tatiana_mamaliga-mt}. In these implementations, we can apply OLAP operations such as \emph{slice}, \emph{dice}, \emph{roll-up}, and \emph{drill-down} on the event data. Assume we are interested in discovering the model of cars produced in the years \emph{2017} and \emph{2018}. As shown in Figure~\ref{fig-cube-slice-dice}, we can slice the cube for the \emph{time} dimension, for the years \emph{2017} and \emph{2018} and find the process model of the remaining events. In fact, through \emph{slicing} the cube, we zoom into a \emph{slice} of the data and remove that dimension from the cube. However, by \emph{dicing}, we apply a filter on multiple dimensions. Assume we are interested in discovering the model for cars with white and blue color produced in years \emph{2017} and \emph{2018}. As shown in Figure~\ref{fig-cube-slice-dice}, we can dice the cube for dimensions \emph{color}, and \emph{time} and discover the model of the remained events. 

\begin{figure}[h!]
\centering
\includegraphics[ width=1\textwidth,height = 0.27\textheight]{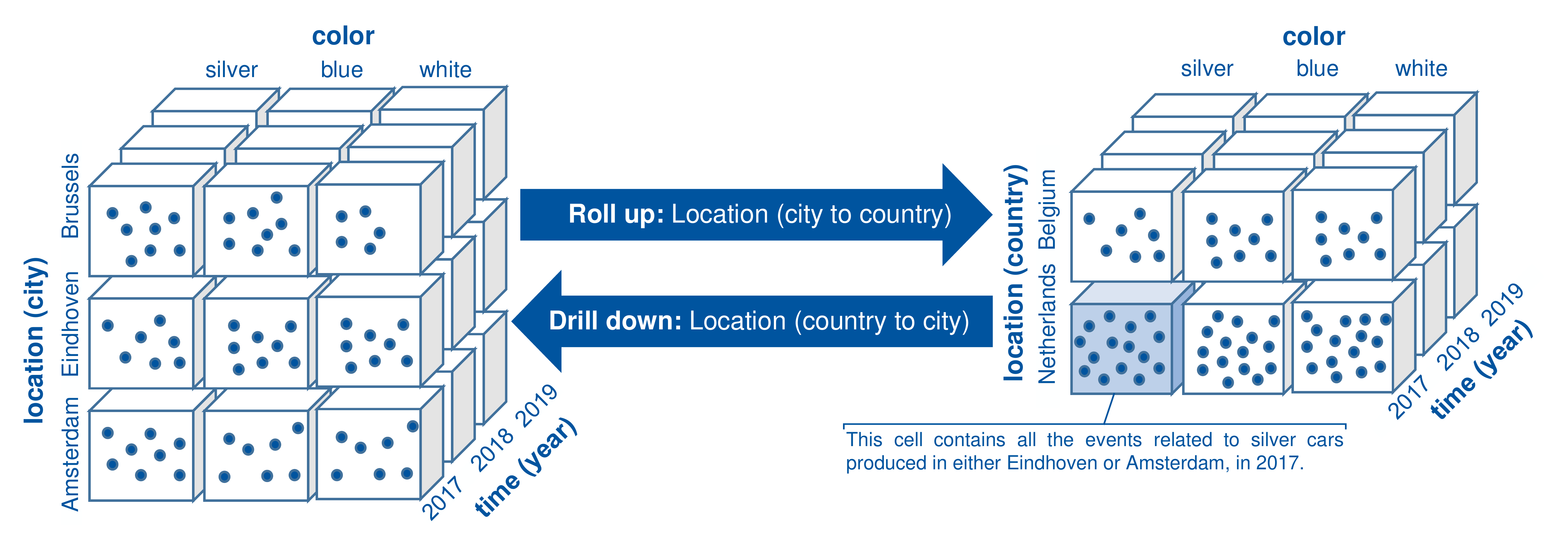}
\caption{Example of \emph{rolling up} and \emph{drilling down} operations in process cubes.}\label{fig-cube-drilldown-rollup}
\end{figure}

\emph{Roll-up} and \emph{drill-down} operations, show the cube with different levels of granularity. An example is shown in Figure~\ref{fig-cube-drilldown-rollup}. We can drill down the \emph{location} dimension from \emph{country} to \emph{city}. \emph{Roll-up} is the opposite of \emph{drill-down} operation. It performs aggregation on a dimension. As shown in Figure~\ref{fig-cube-drilldown-rollup}, we can roll up the \emph{location} dimension from \emph{city} to \emph{country}.

\begin{figure}[t]
\centering
\includegraphics[width=\textwidth]{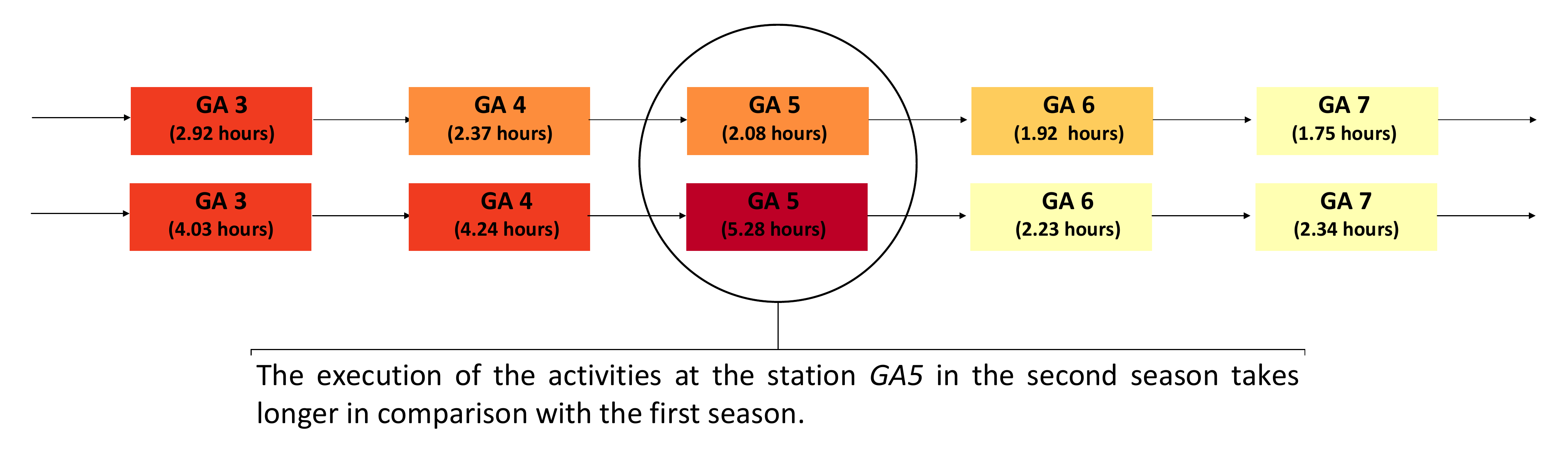}
\caption{Example of comparing the company in two different time windows.}\label{fig-model-comparison}
\end{figure}

A brief example of the application of process cube operations for the generated data is shown in Figure~\ref{fig-model-comparison}. In this figure, the models of two different \emph{slices} corresponding to two different seasons of a year are compared to each other. We can compare performance of the stations with each other through performance analysis. As indicated in the figure, the average time of execution of activities at the station \emph{GA5} has increased in the second season of the year.  

Although there are several implementations of process cubes, some challenges remain. In OLAP, we reduce a set of values into a single summarized value such as the average. However, process cubes deal with event data and we cannot simply reduce many events into one event. The ways that are used for summarizing the events affect the performance of the process cube. Consequently, defining a way for summarizing the events in the process cube with high performance is an open issue.

\subsection{Concept Drift}
\label{sec:drift}

Even though operational processes should ideally be stable, complex internal and external dependencies will typically lead to continuous changes of the process. The resulting variability is, on the one hand, problematic for traditional process mining techniques, which implicitly assume steady behavior. On the other hand, changes can also require an immediate adaptation by the process managers.
In principle, we can attribute the aforementioned changes to external factors, e.g., suppliers, market demand, or political regulations, and internal factors, e.g., operators, minor changes of the \emph{Bill-of-Material} (BOM), or machines. While the former are under limited control of the company, evaluating their effect by means of comparative process mining can support mid- and long-term decision making. In addition, controlling the latter, in particular, has the potential to continuously improve and steer the operational process. 
Therefore, \emph{concept drift detection}, as a field of research that is concerned with detecting systematic changes in processes, is important for monitoring operational processes and providing entry points for further analysis.

A general challenge for methods that systematically assess changes in the execution of a process arises from the different natures of occurrence and duration~\cite{DBLP:conf/caise/BoseAZP11}. For example, \emph{sudden drifts} are caused by rapid process changes and might require instant action while \emph{gradual drifts} describe slowly changing processes where different behaviors overlap. Concerning the long-term development of the process, sequences of multiple changes, i.e., \emph{incremental drifts} and \emph{recurring drifts}, which describe periodical behavior changes, can be distinguished. On the one hand, detecting them is essential in order to understand the overall process evolution. On the other hand, being aware of recurring, e.g., seasonal changes enable process owners to anticipate and adapt while also setting the context for additional analyses.

\begin{figure}[tb]
    \centering
    \includegraphics[width=\linewidth]{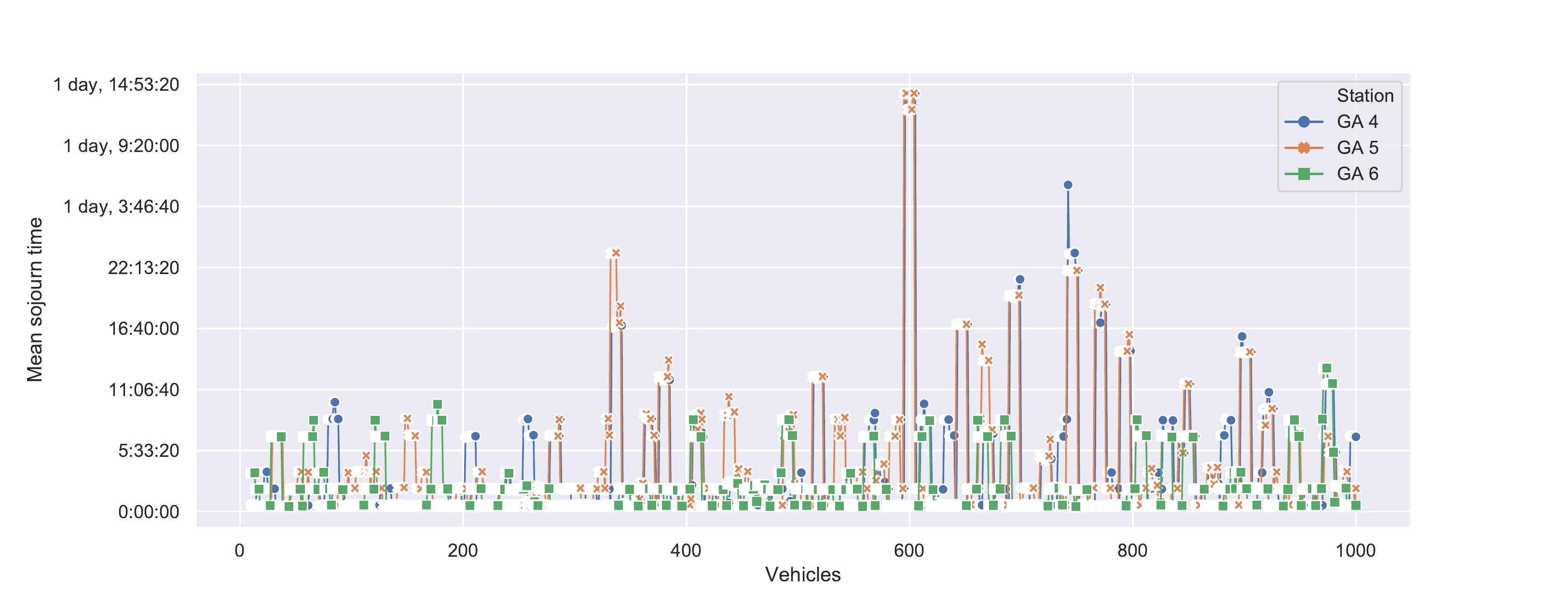}
    \caption{Mean sojourn time for stations \emph{GA4}, \emph{GA5}, and \emph{GA6} over the last ten vehicles. It shows two drifts for \emph{GA4} and \emph{GA5} after 350 and 600 cars, respectively.}\label{fig-drift}
\end{figure}

While existing work focuses on the data~\cite{DBLP:conf/simpda/HompesBADB15} and especially the control-flow perspective~\cite{DBLP:conf/bpm/MaaradjiDRO15,DBLP:conf/caise/BoseAZP11,DBLP:journals/tnn/BoseAZP14}, it neglects time, such as waiting times or activity service times, as a factor of major concern in operational processes. 
As in many operational processes activities are highly structured and tightly coupled, the impact of small delays can accumulate causing a significant decrease in the overall performance. For instance, in our running example delays at sub-assembly stations can temporarily stop significant parts of the entire production line. However, changes such as a battery shortage do not change the ground truth control flow but increase the sojourn times at affected stations making time-aware drift detection essential. 
For example, consider the rolling mean sojourn times over the last ten vehicles for a selected number of stations that are depicted in Figure~\ref{fig-drift}. While it shows two increases for the stations \emph{GA4} and \emph{GA5} after the first 350 and 600 vehicles, respectively, the mean sojourn times for station \emph{GA6} are not affected. This indicates a concept drift that worsens the situation at the bottleneck station \emph{GA5} blocking the preceding stations and, therefore, increasing their sojourn times. Using additional background knowledge, this drift can be attributed to problems with the battery supply and, later on, a severe battery shortage. 
In general, detecting time drift and its causes becomes especially challenging if changes are not local to a station but distributed over the production line. 

In addition to the time perspective, change point detection methods that generally allow for different perspectives on the operational process are essential in order to incorporate domain knowledge, e.g., from the BOM. 
Consequently, this enables the user to focus the sensitivity of the detection method on specific parts of the process while reducing the sensitivity for irrelevant changes. 
For example, the user might focus on changes of important quality metrics and changing impacts of preceding production steps on the former instead of detecting changes 
in operator-dependent activity orderings.
Apart from the change identification, a good visualization showing the nature of the change~\cite{DBLP:conf/caise/BoseAZP11} is crucial to make changes actionable. While change characterization of the control flow has been addressed in~\cite{DBLP:journals/tkdd/OstovarLR20,DBLP:conf/caise/OstovarMRH17}, change characterization of operational processes should acknowledge additional perspectives.

Consequently, in order to address these challenges and unravel the process dynamics, holistic and especially time-aware change point detection approaches with a multi-perspective view on the operational process are needed.

\subsection{Comparison}
\label{sec:comp}

Applying the concepts from Sections~\ref{sec:processcube}~and~\ref{sec:drift}, allows us to organize the data into process cubes, potentially refined by concept drift detection results. Subsequently, this organization provides us a framework and entry points for process comparison along the dimensions of variability which can be challenging even in case of a stable control flow. 
Depending on the inter-activity dependencies, pairwise activity comparison between two process variants can be insufficient and activity relations, i.e., the process knowledge need to be incorporated. 

In order to illustrate the challenges for time-centered process comparison introduced by inter-activity dependencies, we complement the baseline model by a human factor, which expresses in operator-dependent service times, and compare two factories of the same car manufacturing company at different locations. Similar to the workforce assignment in the underlying real-world process of the electric vehicle manufacturer, we divide the production pipeline into multiple sections. Each section has a separate pool of operators from which free operators are assigned to the manufacturing tasks. Moreover, we also model operator preferences for certain stations. Accordingly, when a vehicle arrives at an assembly station, a free operator is randomly assigned based on the current list of preferences in the pool. We generate events for two factories located in the Netherlands and in Belgium and store them in a process cube.

\begin{figure}[tb]
  \begin{subfigure}[c]{\linewidth}
    \centering
    \includegraphics[width=0.5\linewidth]{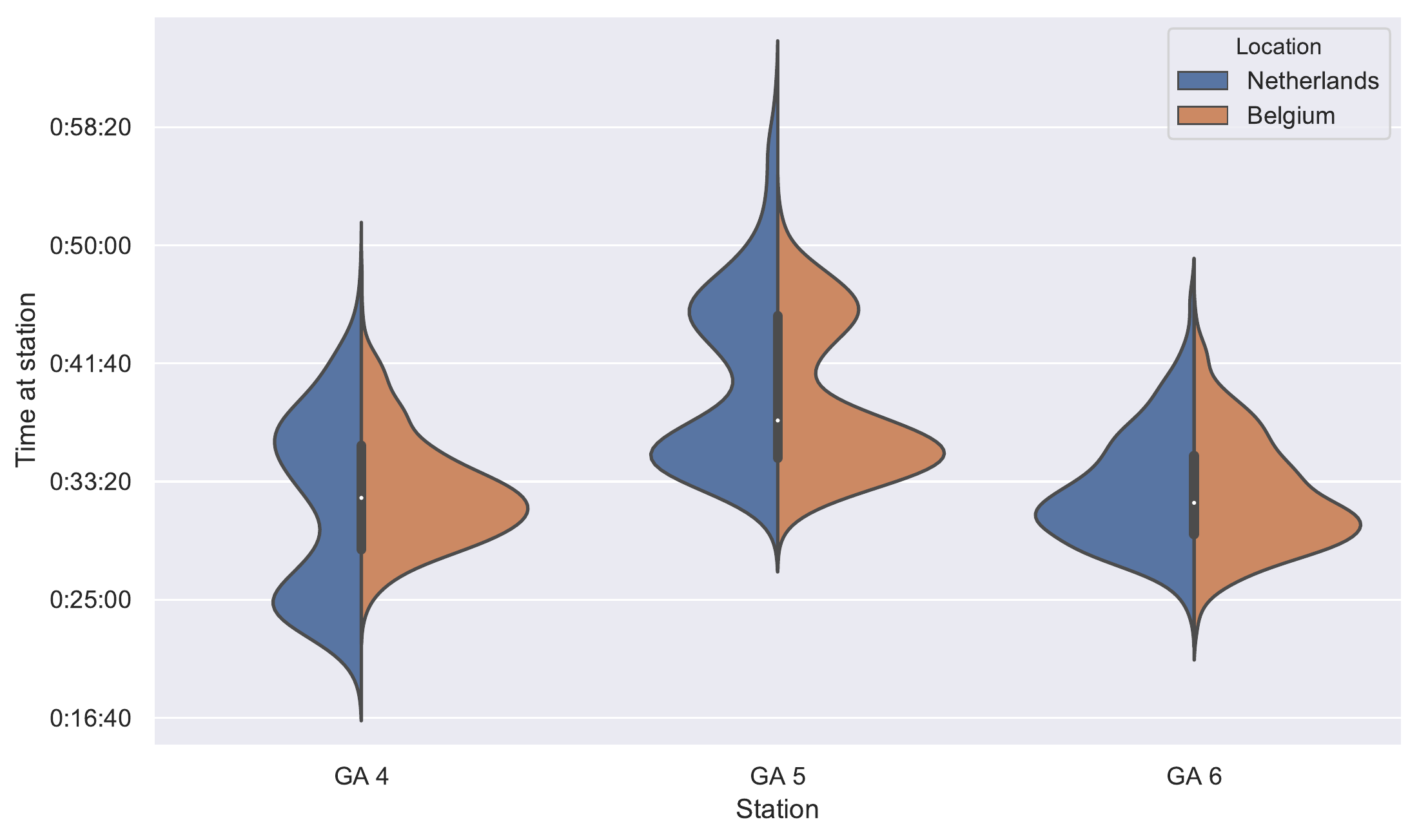}
    \caption{Activity-wise comparison.}\label{fig-comp-actwise}
  \end{subfigure}\\
  \begin{subfigure}[c]{0.48\linewidth}
    \includegraphics[width=\linewidth]{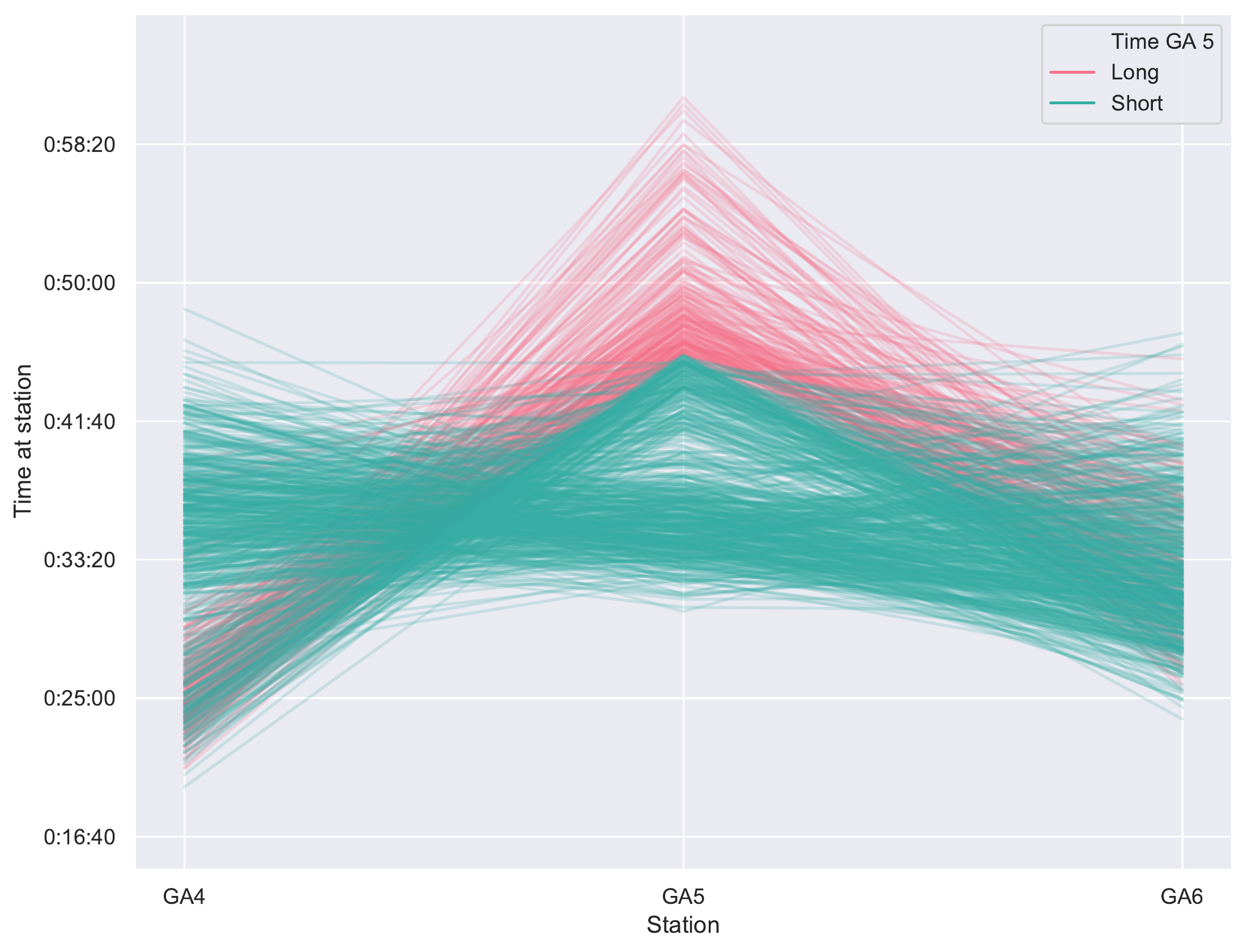}
      \caption{Relations between the station durations in the Netherlands.}\label{fig-comp-proc-Neth}
  \end{subfigure}\hfill%
  \begin{subfigure}[c]{0.48\linewidth}
    \includegraphics[width=\linewidth]{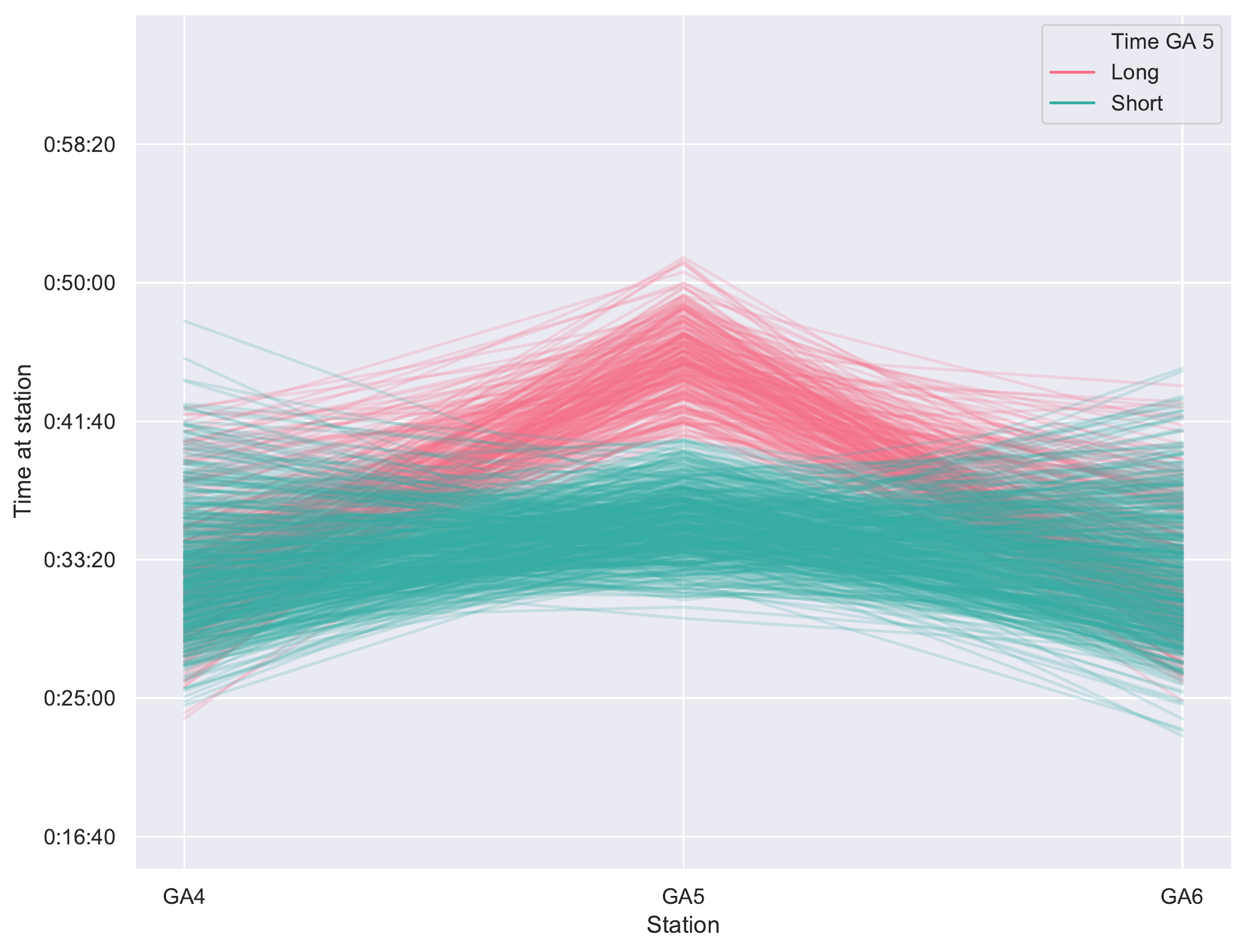}
      \caption{Relations between the station durations in Belgium.}\label{fig-comp-proc-Belg}
  \end{subfigure}
  \caption{A comparative analysis of the times spent in the different stations in both the Netherlands and Belgium.}\label{fig-comp}
\end{figure}

The process owner is interested in comparing the two implementations of his manufacturing process in order to identify possibilities for improvement. To this end, we first \emph{slice} and \emph{dice} the process cube in order to generate a sublog for each location.
As we show next, comparison methods should consider the inter-activity dependencies in order to provide a holistic analysis to the company. Activity-wise comparison, as depicted in Figure~\ref{fig-comp-actwise}, yields that the station duration distributions for the station \emph{GA6} are similar to each other, while the distributions differ for the stations \emph{GA4} and \emph{GA5}. For the station \emph{GA5}, the station duration distribution has a longer tail in the Dutch factory, which indicates inefficiencies for the Dutch location. Considering station \emph{GA4}, the distribution for the Netherlands has two peaks(but not for the location in Belgium) suggesting that it is possible to considerably improve processing times at this station.

However, a process-oriented view, as shown in Figures~\ref{fig-comp-proc-Neth} and~\ref{fig-comp-proc-Belg}, that considers additional context information, yields a different conclusion. The figures show the relations between subsequent stations where traces are colored according to their time at the station \emph{GA5}. Figure~\ref{fig-comp-proc-Neth} shows that the time improvement for the station \emph{GA4} correlates with an increased duration at the station \emph{GA5}. Quick but imprecise work at this station that requires additional rework increases the overall effort. Process comparison reveals this process insight and the absence of this relation in the Belgian factory. Furthermore, the modes of station \emph{GA5} in the Belgian factory seem to be uncorrelated with the time for station \emph{GA4} suggesting the existence of an additional latent variable. Comparison with Figure~\ref{fig-comp-proc-Neth} and the absence of a corresponding pattern in Figure~\ref{fig-comp-proc-Belg}, further suggest that this latent variable is not inherent in the baseline process, but related to the location. For example, different operators working at different speeds could cause this behavior rather than imprecisely fitting parts, causing problems in the next step.

This example shows that, although simple comparison approaches can provide insights into the differences between the processes, they might not consider all the relevant context information.
This motivates the use of flexible comparison methods such as the \emph{Earth Mover's Distance} (EMD) \cite{Wasserstein,DBLP:journals/ijcv/RubnerTG00}. On the one hand, these methods should be sensitive to the frequency of the occurring patterns and thus to their relevance. On the other hand, they should also allow for different perspectives on the process.

By considering the car as the case identifier, we examined the process using general process mining techniques (e.g., process discovery, and conformance checking) and process comparison approaches. However, in reality, the data extracted from the information system of a car factory contains other case identifiers, e.g., orders, and customers. Therefore, in our analysis, we need to provide a more holistic vision of the process by considering multiple case notions, as is discussed next.

\section{Object-Centric Process Mining}
\label{sec:oopm}

Object-centric process mining is an emerging branch of process mining aiming to apply process mining techniques on event logs which are closer to the data extracted from information systems. Multiple objects such as customers, orders, and deliveries are involved in the car factory, described in Section~\ref{sec:re}. Analyzing the processes covering all these aspects is addressed in object-centric process mining. 

\begin{figure}[hthh]
{
\centering
\includegraphics[width=0.85\textwidth, height= 0.19\textheight]{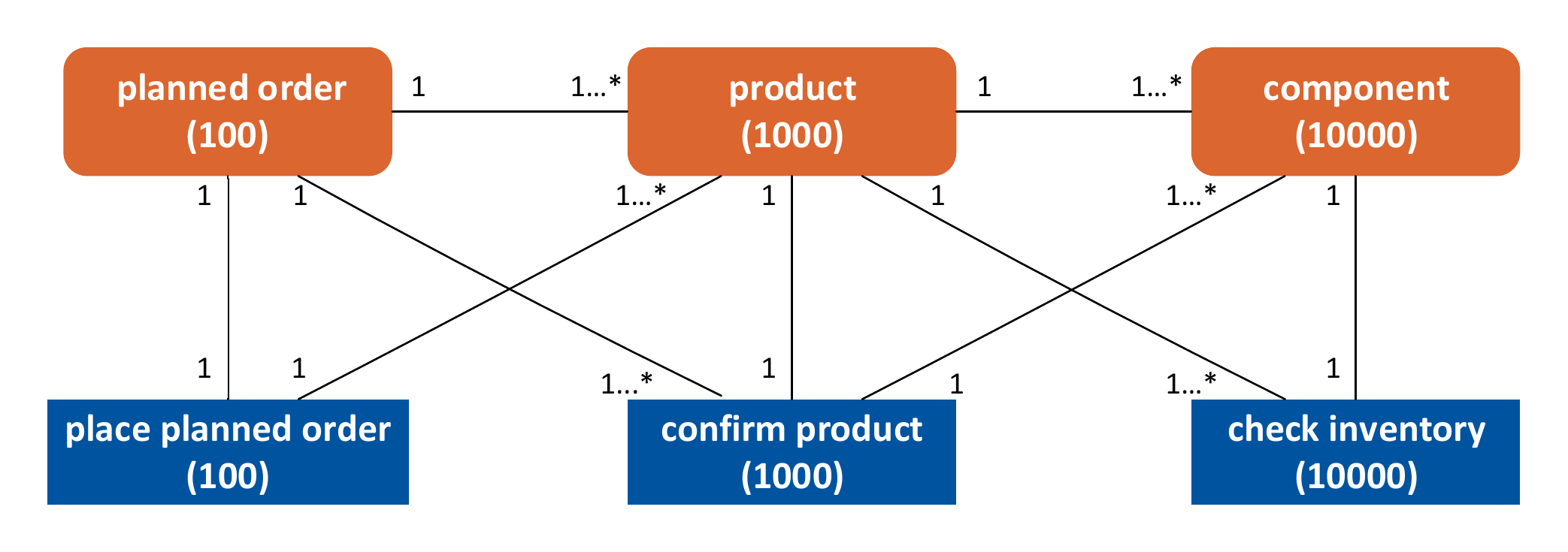}
\caption{Overview of the relationship among case notions (i.e., \emph{planned order, product}, and \emph{component}) and activities (i.e., \emph{place planned order}, \emph{confirm product}, and \emph{check inventory}). For example, there is one-to-one relationship between \emph{planned order} and \emph{place planned order} and one-to-many relationship between \emph{planned order} and \emph{product}.}\label{fig-obj-conv-div}
}
\end{figure}

Companies record their information in information systems such as the ERP (Enterprise Resource Planning) systems of SAP, which do not have the structure of traditional event logs. In traditional event logs, each event refers to a single case notion (i.e., a process instance), an activity, a timestamp, and any number of additional attributes (e.g., costs, resources, etc.). 
However, in operational processes with many interactions (i.e., an event is related to multiple objects), it may be problematic to create and analyze traditional event logs with a single case notion~\cite{DBLP:conf/sefm/Aalst19}. Information systems that support production, such as SAP, store information in the related tables of a database. In the production planning module of SAP, multiple objects (e.g., planned order, supplier, product, component, and delivery) are involved. Each planned order consists of many products, each comprising a range of components (e.g., based on the BOM a car is composed of the frame, engine, battery, etc.). Accordingly, we are able to study the process from multiple different angles and dimensions. Extracted event logs from SAP systems usually suffer from \emph{convergence} 
and \emph{divergence}
~\cite{DBLP:conf/sefm/Aalst19}. These two problems are of high importance to discuss, because they cause challenges in applying process mining techniques, e.g., process discovery, on these event logs. To illustrate these two problems, as an example, consider the simplified process of production planning shown in Figure~\ref{fig-obj-conv-div} with three case notions (i.e., planned order, component, and product) and three activities (i.e., \emph{place order}, \emph{confirm products}, \emph{check the inventory}), respectively. The mentioned problems in this process are:

\begin{itemize}
  \item \emph{Convergence}:
  Events referring to multiple objects of the selected type are replicated, possibly
leading to unintentional duplication \cite{DBLP:conf/sefm/Aalst19}.
For example, an order may contain many products and each product may contain different components. 
Assume that there is an order comprising 10 products, e.g., cars. 
In order to apply classical process mining techniques, we need to flatten the event data by picking a case notion. 
If we select the product as a case notion, 
then this leads to 10 replications of the same \emph{place order} event. 
The duplication of events for different cases may lead to an explosion of the number of events. 
Moreover, when time and costs are associated to events, this may lead to very misleading insights
In object-centric event logs, an event may contain references to many objects of the same case notion, thus avoiding the duplication problem.
  \item \emph{Divergence}:
Events referring to different objects of a type not selected as the case notion are
considered to be causally related but are executed independently \cite{DBLP:conf/sefm/Aalst19}.
Assume that we pick order as a case notion (to avoid convergence problems).
A single order may contain multiple products and each product may contain different components. 
However, there are activities executed for a single product or a single component, e.g., \emph{check inventory}.
Many \emph{check inventory} events may refer to the same order although they actually independent and refer to different objects. 
Since we picked order as a case notion, these events cannot be disentangled, typically leading to Spaghetti models. 
Things that happen in a strict order for both components and products become blurred when using a coarser case notion.
\end{itemize}

Object-centric process mining aims to provide a solution for the aforementioned challenges.
The interest in this subdiscipline is rapidly increasing, because organizations are in need of a more holistic way to interact with event logs extracted from information systems~\cite{DBLP:journals/corr/abs-2001-02562,DBLP:journals/kais/MurillasRA20,DBLP:conf/caise/LiMCA18}.
 Several techniques were developed to deal with object-centric event logs for process analysis:
\begin{itemize}
  \item Extracting object-centric event logs from information systems: This includes several contributions related to the storage format and some work on the extraction from SAP logs or ERP systems in general~\cite{DBLP:journals/corr/abs-2001-02562,DBLP:journals/kais/MurillasRA20,simovic2018domain}.
  \item Discovering process models from object-centric event logs: Artifact-centric modeling is an approach to model processes with multiple case notions by combining process and data~\cite{DBLP:journals/debu/CohnH09,DBLP:conf/IEEEscc/NarendraBTM09}. The techniques proposed based on the artifact-centric process modeling do not show the process as a whole. Therefore, in~\cite{DBLP:conf/sac/LiCA19} a discovery algorithm was proposed to discover Object-Centric Behavioral Constraints (OCBC) models from object-centric event logs. These models show interactions between the data and behavioral perspectives on the attribute level in one diagram. The main challenge of OCBC is scalability and complexity, which led to the development of MVP (Multiple Viewpoint) Models. MVP models are Directly-Follows Graphs (DFG) with colored arcs annotated by frequency and performance information. MVP models show the process model without omitting any of the case notions~\cite{DBLP:journals/corr/abs-2001-02562}. 
MVP models cannot capture concurrency well and the diagnostics may be misleading. 
This led to the development of techniques to discovered colored Petri nets.  
Object-centric Petri nets are another type of object-centric process models that can be extracted from object-centric event logs and provide the execution semantics~\cite{ocpn_fi_2020}.\end{itemize}

\begin{figure}[tb]
{
\centering
\includegraphics[width=0.70\textwidth, height= 0.40\textheight]{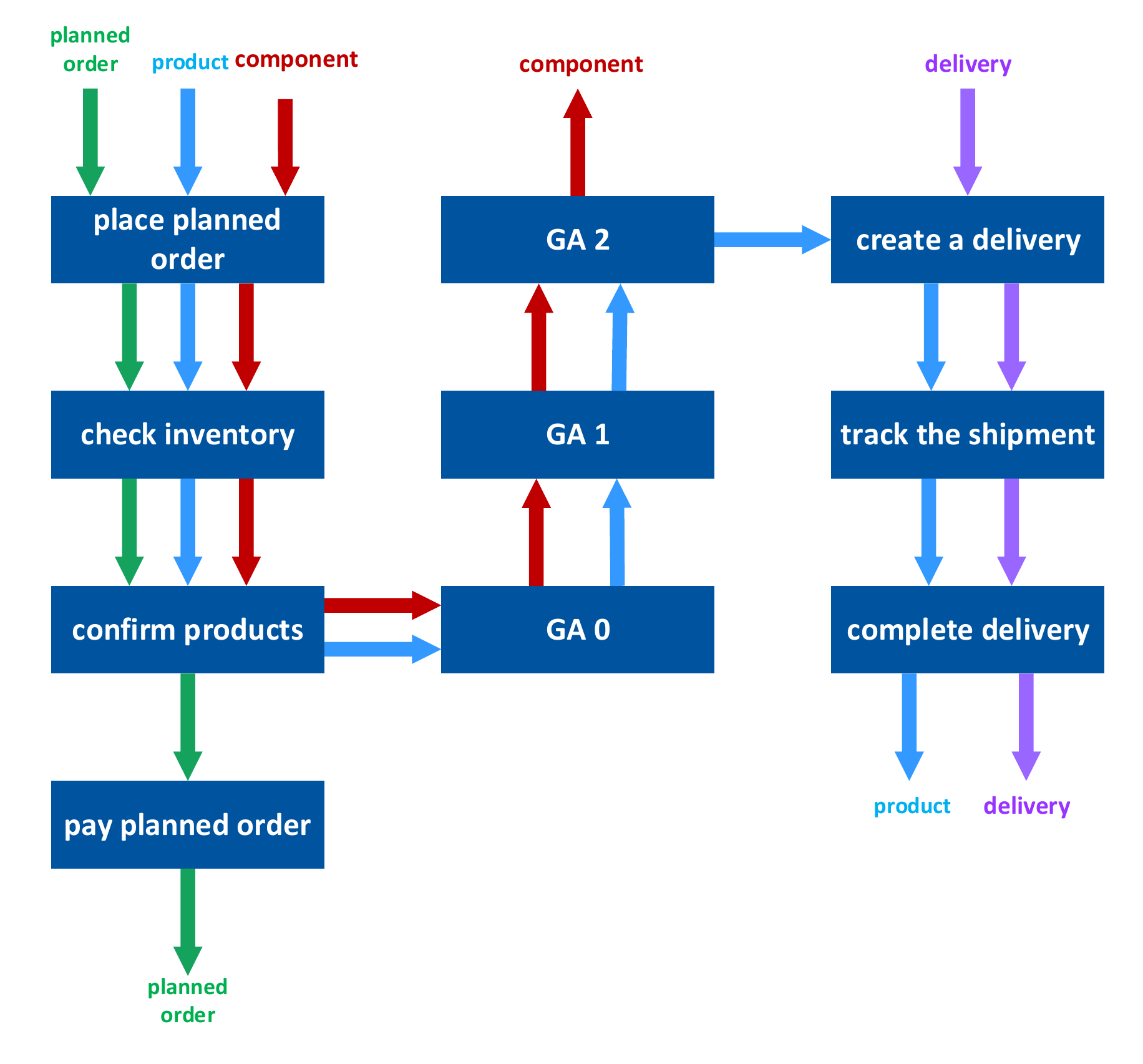}
\caption{Directly-Follows Multigraph for the case notions order, product, component, and delivery. A fragment of the production line is captured that consists of the stations \emph{GA1}, \emph{GA2}, and \emph{GA3}. }
\label{fig-obj-model}
}
\end{figure}

A baseline discovery model for object-centric processes is the Directly-Follows Multigraph, inspired by DFGs~\cite{DBLP:conf/sefm/Aalst19}. DFGs are graphs where the nodes represent activities or events in the event log. Nodes are connected through edges if there is at least one case in the event log where the source event or activity is followed by the target event or activity~\cite{DBLP:books/sp/Aalst16}. For generating Directly-Follows Multigraph, we merge the DFGs that are generated for each case notion into an overall DFG where the arcs with different colors correspond to different case notions. In Figure~\ref{fig-obj-model}, a Directly-Follows Multigraph for the extended model of the car factory with multiple case notions (e.g., planned order, products, component, and delivery), is shown. Following the green arcs, the order goes through \emph{place planned order}, \emph{check inventory}, \emph{confirm products}, and \emph{pay order}. The product and component enter the production line which consists of the stations \emph{GA0}, \emph{GA1}, and \emph{GA2}. The component leaves the production line at the last station (\emph{GA2}), while the last step for any product is \emph{complete delivery}.

Using object-centric process models such as Directly-Follows Multigraph shown in Figure~\ref{fig-obj-model}, we gain insights into object-centric processes. However, techniques developed for object-centric process mining can still focus on single location, time period, or process variant. There may exist multiple variants of the same process, which motivates the use of process comparison techniques such as process cubes, supporting the analysis of object-centric event logs. To adapt the concept of process cube operations to object-centric process mining is not trivial since an event may refer to any number of objects. \emph{Slicing} for dimensions related to case notions may lead to convergence and divergence problems. For example, events related to a specific component also contain other components. \emph{Dicing} suffers from the same problem. Therefore, traditional process cubes cannot handle object-centric event logs and it is worthwhile to bridge the gap between process comparison approaches and object-centric process mining.
This needed to compare the processes in a more holistic setting.


\section{Forward-Looking Process Mining}
\label{sec:fpm}

The value of process mining in analyzing the past is clear and widely accepted.
However, just diagnosing the past is not a goal in itself.
The actual goal is to continuously improve processes and respond to changes. 
Operational processes are subject to many changes, e.g., a sudden increase in the number of orders, and therefore the managers require an extended vision of the future in order to deal with changes in the process.
Due to the high cost of operational processes, in the face of deliberate changes in order to improve the operational processes performance or unexpected changes, having the ability to look forward is of paramount importance. 
Simulation is capable of enabling process mining to look forward.
At the same time, data-driven support provided by process mining, e.g., past executions of the process and process model, can make the simulation models more realistic.
In order to answer forward-looking questions regarding the future of processes using process mining, the first step is to demonstrate how process mining and simulation can complement each other.
The knowledge gained about the process (including the discovered process model and the current performance of the process) can be used to predict future states of the running cases (e.g., the remaining flow time) under the assumption of a stable process, i.e., no changes in the process. 
However, process mining is backward looking and cannot be used for ``What if?'' questions. 
Hence, process mining and simulation perfectly complement each other~\cite{DBLP:conf/scsc/Aalst18}.
Among the different simulation techniques, we consider two different approaches: discrete event simulation or system dynamics simulation. These techniques are both able to model the operational processes at different levels of detail by capturing process events and variables.

\subsection{Discrete Event Simulation}
\label{sec:des}

One of the well-known simulation techniques which can be used for simulating operational processes is Discrete Event Simulation (DES). 
DES uses predefined rules according to which the simulation process generates events.
Whenever an event occurs, the state of the system changes and the new state is recorded.
Each state enables new events. 
Each possible way is described by a simulation run which shows the result of the play-out of the model~\cite{DBLP:conf/scsc/Aalst18}. 
Tools such as Coloured Petri Nets (CPNs) can be used for simulating a process including all the details of a process, e.g., number of resources for each activity and duration of each activity~\cite{CPNtoolsEditing}.
Predefined key performance indicators such as average waiting time for each process instance can be calculated and compared in different situations by redesigning the simulation models and experiments~\cite{DBLP:conf/scsc/Aalst18}. 
Despite the capable simulation tools along with the provided forward-looking approaches, the real-life application of simulation is limited~\cite{DBLP:series/ihis/Aalst15}.
Detailed simulation models like DES may be very time-consuming to build. Interpreting simulation results is not an easy task and 
often models need to be tuned to behave similar to the real process 
(coffee breaks, visits to the toilet, holidays, illness, etc.\ lead to lower performance than expected based on the initial simulation model).
Furthermore, simulation models tend to capture only some aspects of a process or use an oversimplified model of reality~\cite{DBLP:conf/caise/Aalst10}.
Therefore, organizations try to use the evidence-based approaches in which the previously captured state of the organization can be used for more accurate simulation models~\cite{DBLP:conf/scsc/Aalst18}.
In order to achieve more realistic results using simulation techniques, different approaches have been proposed~\cite{disc_sim_models_is,sim-YAWL-ProM-dke}. 

Discrete Event Simulation (DES) requires considerable domain knowledge of the process.
For instance, to answer the forward-looking question for our example production line, we need to implement all the details of the production line to see the effect of inventory options for one of the stations on the overall production line. 
Also, it is not easy to consider the context in which the simulation is running~\cite{process-mining-put-into-context2011}. 
Incorporating effects of external factors in the models, e.g., human behaviors
or environmental variables are other aspects that are missing from current simulation techniques in processes~\cite{DBLP:conf/otm/PourbafraniZA19}. 
Moreover, selecting the right level of abstraction and avoiding too much detail are other important aspects of creating a simulation model and performing simulation. 
Unfortunately, the inherently required level of detail in these types of approaches does not allow for high-level modeling and long-term predictions.
As opposed to the existing simulation techniques, System Dynamics (SD) allows us to assess the impact of changes in the process from a global perspective as well as the effects of external factors.
Using different levels of granularity in the modeling, we can address
the major drawbacks of discrete event simulation techniques.
Therefore, the goal is to assess and predict the future behavior of operational processes including the effects of potential changes as well as the roles of external factors, e.g., human behavior at an aggregated level. 

\subsection{System Dynamics}
\label{sec:sd}
System Dynamics (SD) techniques are used to model dynamic systems and their relations with their environment. 
Using system dynamics, the factors affecting a system’s behavior are captured~\cite{sterman2002system}. 
Stocks, flows and variables are fundamental elements in modeling a system as a stock-flow diagram, i.e., one of the main system dynamics diagrams. 
\begin{figure}[tb]
{
\centering
\includegraphics[width=0.45\textwidth]{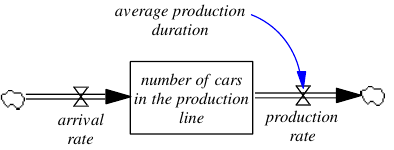}
\caption{A sample stock-flow diagram for the running example. The value of the stock \emph{number of cars in the production line} is calculated from the value of \emph{arrival rate} and \emph{production rate} as flows per time step, e.g., per day. Also, the value of the \emph{production rate} is affected by the value of \emph{average production duration}.}
\label{fig-sd-sample-sfd}
}
\end{figure}
Figure~\ref{fig-sd-sample-sfd} shows a simple stock-flow diagram for the running example in which \emph{arrival rate} and \emph{production rate} as flows add/remove to/from the values of \emph{number of cars in the production line} as stock, also, \emph{average production duration} (a variable) affects \emph{production rate}.
System dynamics provide the opportunity to add exogenous factors, e.g., variables in stock-flow diagrams as well to capture the hidden relationships among the players in the operational processes~\cite{DBLP:conf/otm/PourbafraniZA19}.
As shown in Figure~\ref{fig-pm-types}, different executions of operational processes that are stored in the information systems can be captured in the form of event logs. 
The provided event logs for operational processes can be transformed in other formats to be used as inputs for simulation techniques.
These transformed event logs include the performance metrics of the processes over different time steps, e.g., arrival rate, average service time and the number of waiting items over steps of time in the process~\cite{mahsaBIS2020}.

\begin{figure}[tb]
{
\centering
\includegraphics[width=0.9\textwidth,height=0.18\textheight]{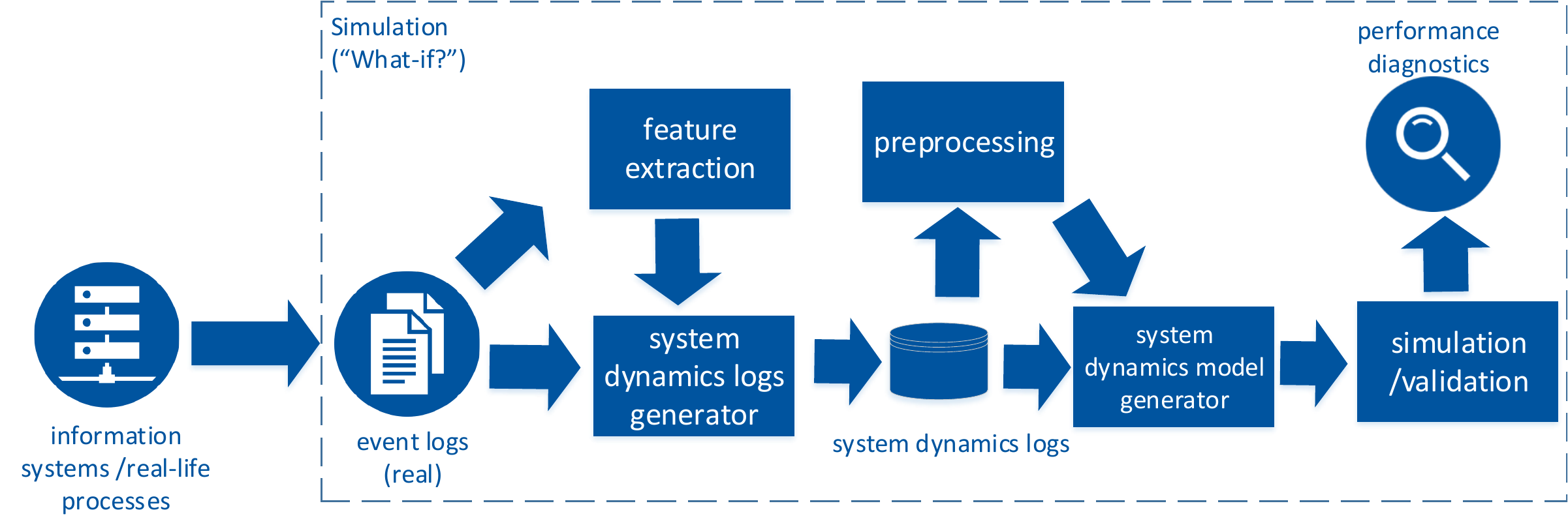}
\caption{A general framework for using process mining and simulation for improving operational processes. Using event logs and transforming them to System Dynamics logs (SD-logs) along investigating the relations between different parameters in the process over time, the simulation regarding answering ``What if?'' questions is possible.}
\label{fig-sd-approach}
}
\end{figure}

The authors in~\cite{DBLP:conf/ihsi/PourbafraniZA20} exploit the general proposed approach in~\cite{DBLP:conf/otm/PourbafraniZA19}, i.e., a simulation approach for the business process using system dynamics and process mining techniques, which is designed for the specific purpose of simulating the production lines w.r.t performance metrics.
Therefore, as shown in Figure~\ref{fig-sd-approach}, for operational processes, the simulation model in the form of a system dynamics model using trained variables based on event logs can be generated.  
We use system dynamics and process mining together to answer ``What if?'' questions, e.g., what will happen if we add one more resource to one of the stations? 
We transform the generated event log into a \emph{System Dynamics log} (SD-Log) which describes the values of a collection of variables of the process over time using the time window selection technique in \cite{Mahsatimeseries}. Then, we use the calculated values of the variables to detect the relationships between metrics and generating a stock-flow diagram which is used for simulation~\cite{mahsaBIS2020}. 
\begin{figure}[tb]
{
\centering
\includegraphics[width=0.70\textwidth,height=0.2\textheight]{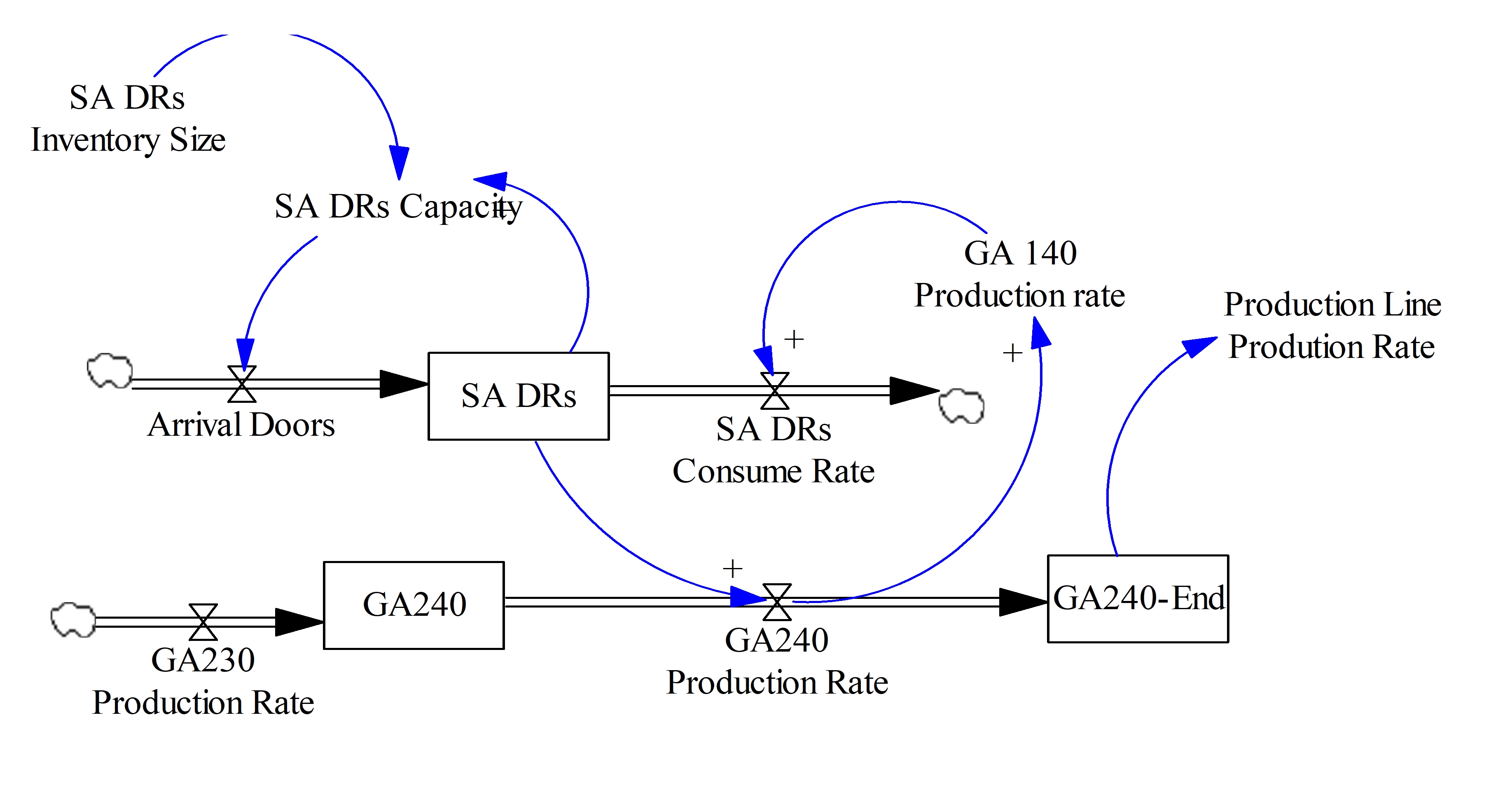}
\caption{A sample stock-flow diagram generated based on the production line event log. The purpose of the model is to simulate the effect of adding an inventory option to the doors sub-assembly on the final production rate of the production line at an aggregated level, e.g., per month. }\label{fig-sd-evaluation}
}
\end{figure}
\begin{figure}[tb]
{
\centering
\includegraphics[width=0.75\textwidth, height=0.25\textheight]{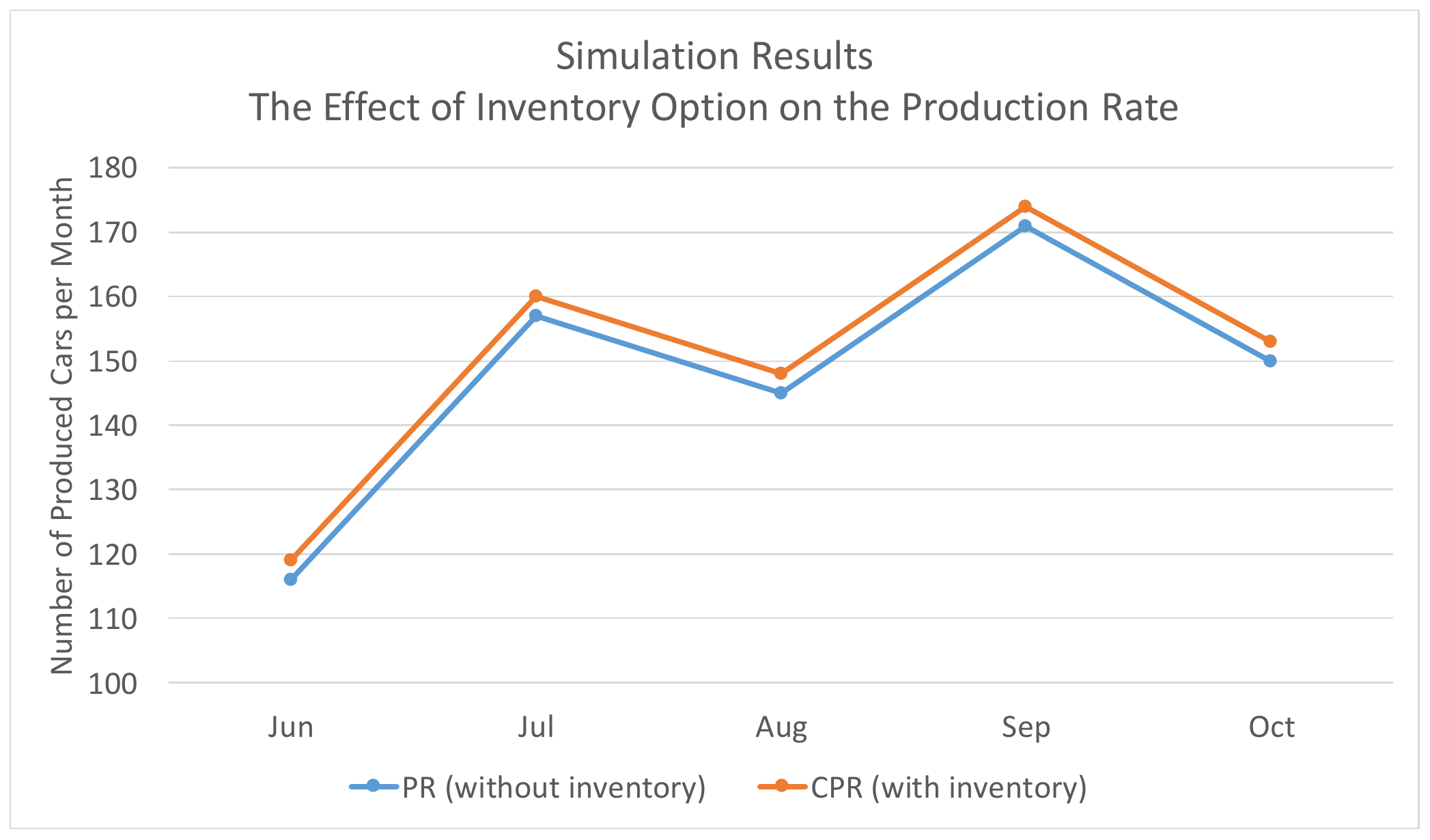}
\caption{A comparison between the actual production rate (PR) and the simulated PR using our approach along actually changed production rate (CPR) and simulated CPR for adding an inventory to the doors sub-assembly, e.g., in August 3 more cars will be produced.}\label{fig-sd-evaluation-res}
}
\end{figure}

\subsection{Forward-Looking in Production Lines}
As indicated in Section~\ref{sec:re}, the stations in the base model do not have an inventory or queuing option.
As a simplified example, consider the station \emph{GA24}, at which the cars at this station require the doors prepared in \emph{SA7}, i.e., a door preparation sub-assembly station.
Therefore, in the production line, there can be situations in which a car is at the station \emph{GA24} and cannot leave the station because it is waiting for the doors to be prepared. 
One of the scenarios to improve the above-mentioned situation is to prepare the doors for each car beforehand and keep them at the sub-assembly station.
The possibility of having a temporary queue, especially for sub-assembly stations, is one of the possible actions which potentially leads to an improved production rate of the whole production line. 
The question to be answered is: How does the temporary queue for the door assembly stations affect the overall production line?
Using the  approach presented in \cite{mahsaToolPMSD}, we can measure the effect of adding an inventory option to only one of the sub-assembly stations at the production rate of the whole production line. We do not need to consider the details required by DES approaches. 

We also populate the generated simulation model, i.e., the stock-flow diagram, with the values generated from the event log of the current status of the production line. 
We use the data generated in a regular situation in which there is no inventory option for the sub-assembly station, 
and after transforming it to the system dynamic log, we generate the model using the proposed approach in Figure~\ref{fig-sd-approach}. 
The result of simulation after adding the inventory option is shown in Figure~\ref{fig-sd-evaluation-res}. 
As expected, the whole production rate has increased and the production line manager can decide whether to invest 
in adding the possibility to keep inventory for the door preparation sub-assembly station. 
The described scenario shows the potential of using a combination of process mining and SD-based simulation in order to improve operational processes. 
Using process mining and simulation techniques we are able to answer ``What if?'' questions in the operational processes.

\section{Conclusion}
\label{sec:concl}

In this paper, we discussed how process mining can be used to remove \emph{operational friction} (e.g., rework, delays, waste, and deviations). We discussed the capabilities provided by existing tools and highlighted some of the main challenges addressed in the DFG-funded Cluster of Excellence ``Internet of Production'' (IoP). Process mining is widely used to improve standard processes like the Order-to-Cash (O2C) and the Purchase-to-Pay (P2P). However, in other areas (e.g., in production, materials handling, and logistics) process mining is not yet widely used. We believe that by addressing the challenges discussed in this paper, adoption can be accelerated.

We highlighted three important research directions: comparative process mining, object-centric process mining, and forward-looking process mining.

\emph{Comparative process mining} aims to visualize and show root causes for differences over time or between different organizational units (e.g., two factories producing the same or similar products).
It is valuable to identify differences and generate insights. These may help to detect problems early (e.g., using concept-drift detection) and establish so-called ``best practices'' (i.e., routines that are considered to be correct or most effective).

\emph{Object-centric process mining} addresses one of the main limitations of today's process mining tools. 
In many applications it is not reasonable to assume a single case notion.
Consider for example the Bill of Materials (BOM) in an assembly process.
A single car has about 2000 parts (or even up to 50,000 parts, counting every part down to the smallest screws). 
When an engine is produced it may not be clear in what type of car it will end up. Both the car and the engine have unique identifiers are are related somewhere in the process. This example shows that assuming a single case notion automatically leads to a particular view on the process. Such views are valuable, but at the same time incomplete and potentially misleading.
Object-centric process mining can be used to provide a more holistic view, also avoiding the convergence and divergence problems described.

\emph{Forward-looking process mining} techniques aim to not only diagnose performance and compliance problems in the past,
but to predict and change processes. We are particularly interested in answering ``What if?'' questions. 
Therefore, simulation plays an important role. 
Next to traditional Discrete Event Simulation (DES), we also proposed the use of System Dynamics (SD).
SD tools simulate the process at a higher abstraction level (i.e., the steps are days or weeks rather than individual events). To learn SD models we need to convert traditional event logs to SD-Logs. 
This is far from trivial, but allows extending the model with contextual factors that are not directly related to individual events.

We strongly believe that progress in comparative process mining, object-centric process mining, 
and forward-looking process mining will assist in the creation of realistic \emph{digital shadows} that can be used to manage, control, and improve operational processes. 
Digital shadows based on process mining include both process models and actual event data.
The combination of process models and actual event data allows us to drill-down into the actual process when a problem emerges, predict the trajectory of current process instances, and anticipate the effects of interventions (e.g., adding resources or reordering production steps).

\subsubsection*{Acknowledgments}  Funded by the Deutsche Forschungsgemeinschaft (DFG, German Research Foundation) 
under Germany's Excellence Strategy–EXC-2023 Internet of Production – 390621612. 
We also thank the Alexander von Humboldt (AvH) Stiftung for supporting our research.

\bibliographystyle{splncs04_adapt}
\bibliography{lit_filtered}
\end{document}